\catcode`\@=11					



\font\fiverm=cmr5				
\font\fivemi=cmmi5				
\font\fivesy=cmsy5				
\font\fivebf=cmbx5				

\skewchar\fivemi='177
\skewchar\fivesy='60


\font\sixrm=cmr6				
\font\sixi=cmmi6				
\font\sixsy=cmsy6				
\font\sixbf=cmbx6				

\skewchar\sixi='177
\skewchar\sixsy='60


\font\sevenrm=cmr7				
\font\seveni=cmmi7				
\font\sevensy=cmsy7				
\font\sevenit=cmti7				
\font\sevenbf=cmbx7				

\skewchar\seveni='177
\skewchar\sevensy='60


\font\eightrm=cmr8				
\font\eighti=cmmi8				
\font\eightsy=cmsy8				
\font\eightit=cmti8				
\font\eightbf=cmbx8				

\skewchar\eighti='177
\skewchar\eightsy='60


\font\ninei=cmmi9
\font\ninesy=cmsy9

\skewchar\ninei='177
\skewchar\ninesy='60


\font\tenrm=cmr10				
\font\teni=cmmi10				
\font\tensy=cmsy10				
\font\tenex=cmex10				
\font\tenit=cmti10				
\font\tensl=cmsl10				
\font\tenbf=cmbx10				
\font\tentt=cmtt10				
\font\tenss=cmss10				
\font\tensc=cmcsc10				
\font\tenbi=cmmib10				

\skewchar\teni='177
\skewchar\tenbi='177
\skewchar\tensy='60

\def\tenpoint{\ifmmode\err@badsizechange\else
	\textfont0=\tenrm \scriptfont0=\sevenrm \scriptscriptfont0=\fiverm
	\textfont1=\teni  \scriptfont1=\seveni  \scriptscriptfont1=\fivemi
	\textfont2=\tensy \scriptfont2=\sevensy \scriptscriptfont2=\fivesy
	\textfont3=\tenex \scriptfont3=\tenex   \scriptscriptfont3=\tenex
	\textfont4=\tenit \scriptfont4=\sevenit \scriptscriptfont4=\sevenit
	\textfont5=\tensl
	\textfont6=\tenbf \scriptfont6=\sevenbf \scriptscriptfont6=\fivebf
	\textfont7=\tentt
	\textfont8=\tenbi \scriptfont8=\seveni  \scriptscriptfont8=\fivemi
	\def\rm{\tenrm\fam=0 }%
	\def\it{\tenit\fam=4 }%
	\def\sl{\tensl\fam=5 }%
	\def\bf{\tenbf\fam=6 }%
	\def\tt{\tentt\fam=7 }%
	\def\ss{\tenss}%
	\def\sc{\tensc}%
	\def\bmit{\fam=8 }%
	\rm\setparameters\setbaselines\fi}


\font\twelverm=cmr12				
\font\twelvei=cmmi12				
\font\twelvesy=cmsy10	scaled\magstep1		
\font\twelveex=cmex10	scaled\magstep1		
\font\twelveit=cmti12				
\font\twelvesl=cmsl12				
\font\twelvebf=cmbx12				
\font\twelvett=cmtt12				
\font\twelvess=cmss12				
\font\twelvesc=cmcsc10	scaled\magstep1		
\font\twelvebi=cmmib10	scaled\magstep1		

\skewchar\twelvei='177
\skewchar\twelvebi='177
\skewchar\twelvesy='60

\def\twelvepoint{\ifmmode\err@badsizechange\else
	\textfont0=\twelverm \scriptfont0=\eightrm \scriptscriptfont0=\sixrm
	\textfont1=\twelvei  \scriptfont1=\eighti  \scriptscriptfont1=\sixi
	\textfont2=\twelvesy \scriptfont2=\eightsy \scriptscriptfont2=\sixsy
	\textfont3=\twelveex \scriptfont3=\tenex   \scriptscriptfont3=\tenex
	\textfont4=\twelveit \scriptfont4=\eightit \scriptscriptfont4=\sevenit
	\textfont5=\twelvesl
	\textfont6=\twelvebf \scriptfont6=\eightbf \scriptscriptfont6=\sixbf
	\textfont7=\twelvett
	\textfont8=\twelvebi \scriptfont8=\eighti  \scriptscriptfont8=\sixi
	\def\rm{\twelverm\fam=0 }%
	\def\it{\twelveit\fam=4 }%
	\def\sl{\twelvesl\fam=5 }%
	\def\bf{\twelvebf\fam=6 }%
	\def\tt{\twelvett\fam=7 }%
	\def\ss{\twelvess}%
	\def\sc{\twelvesc}%
	\def\bmit{\fam=8 }%
	\rm\setparameters\setbaselines\fi}


\font\fourteenrm=cmr12	scaled\magstep1		
\font\fourteeni=cmmi12	scaled\magstep1		
\font\fourteensy=cmsy10	scaled\magstep2		
\font\fourteenex=cmex10	scaled\magstep2		
\font\fourteenit=cmti12	scaled\magstep1		
\font\fourteensl=cmsl12	scaled\magstep1		
\font\fourteenbf=cmbx12	scaled\magstep1		
\font\fourteentt=cmtt12	scaled\magstep1		
\font\fourteenss=cmss12	scaled\magstep1		
\font\fourteensc=cmcsc10 scaled\magstep2	
\font\fourteenbi=cmmib10 scaled\magstep2	

\skewchar\fourteeni='177
\skewchar\fourteenbi='177
\skewchar\fourteensy='60

\def\fourteenpoint{\ifmmode\err@badsizechange\else
	\textfont0=\fourteenrm \scriptfont0=\tenrm \scriptscriptfont0=\sevenrm
	\textfont1=\fourteeni  \scriptfont1=\teni  \scriptscriptfont1=\seveni
	\textfont2=\fourteensy \scriptfont2=\tensy \scriptscriptfont2=\sevensy
	\textfont3=\fourteenex \scriptfont3=\tenex \scriptscriptfont3=\tenex
	\textfont4=\fourteenit \scriptfont4=\tenit \scriptscriptfont4=\sevenit
	\textfont5=\fourteensl
	\textfont6=\fourteenbf \scriptfont6=\tenbf \scriptscriptfont6=\sevenbf
	\textfont7=\fourteentt
	\textfont8=\fourteenbi \scriptfont8=\tenbi \scriptscriptfont8=\seveni
	\def\rm{\fourteenrm\fam=0 }%
	\def\it{\fourteenit\fam=4 }%
	\def\sl{\fourteensl\fam=5 }%
	\def\bf{\fourteenbf\fam=6 }%
	\def\tt{\fourteentt\fam=7}%
	\def\ss{\fourteenss}%
	\def\sc{\fourteensc}%
	\def\bmit{\fam=8 }%
	\rm\setparameters\setbaselines\fi}


\font\seventeenrm=cmr10 scaled\magstep3		


\newdimen\rp@
\newcount\@basestretchnum
\newskip\@baseskip
\newskip\headskip
\newskip\footskip


\def\setparameters{\rp@=.1em
	\headskip=24\rp@
	\footskip=\headskip
	\delimitershortfall=5\rp@
	\nulldelimiterspace=1.2\rp@
	\scriptspace=0.5\rp@
	\abovedisplayskip=10\rp@ plus3\rp@ minus5\rp@
	\belowdisplayskip=10\rp@ plus3\rp@ minus5\rp@
	\abovedisplayshortskip=5\rp@ plus2\rp@ minus4\rp@
	\belowdisplayshortskip=10\rp@ plus3\rp@ minus5\rp@
	\normallineskip=\rp@
	\lineskip=\normallineskip
	\normallineskiplimit=0pt
	\lineskiplimit=\normallineskiplimit
	\jot=3\rp@
	\setbox0=\hbox{\the\textfont3 B}\p@renwd=\wd0
	\skip\footins=12\rp@ plus3\rp@ minus3\rp@
	\skip\topins=0pt plus0pt minus0pt}


\def\setbaselines{\maxdepth=4\rp@\baselinestretch=\@basestretchnum}


\def\baselinestretch{\afterassignment\@basestretch\@basestretchnum}
\def\@basestretch{%
	\@baseskip=12\rp@ \divide\@baseskip by1000
	\normalbaselineskip=\@basestretchnum\@baseskip
	\baselineskip=\normalbaselineskip
	\bigskipamount=\the\baselineskip
		plus.25\baselineskip minus.25\baselineskip
	\medskipamount=.5\baselineskip
		plus.125\baselineskip minus.125\baselineskip
	\smallskipamount=.25\baselineskip
		plus.0625\baselineskip minus.0625\baselineskip
	\setbox\strutbox=\hbox{\vrule height.708\baselineskip
		depth.292\baselineskip width0pt }}



\def\makeheadline{\vbox to0pt{\baselinestretch=1000
	\vskip-\headskip \vskip1.5pt
	\line{\vbox to\ht\strutbox{}\the\headline}\vss}\nointerlineskip}

\def\makefootline{\baselineskip=\footskip\line{\the\footline}}

\def\big#1{{\hbox{$\left#1\vbox to8.5\rp@ {}\right.\n@space$}}}
\def\Big#1{{\hbox{$\left#1\vbox to11.5\rp@ {}\right.\n@space$}}}
\def\bigg#1{{\hbox{$\left#1\vbox to14.5\rp@ {}\right.\n@space$}}}
\def\Bigg#1{{\hbox{$\left#1\vbox to17.5\rp@ {}\right.\n@space$}}}


\mathchardef\alpha="710B
\mathchardef\beta="710C
\mathchardef\gamma="710D
\mathchardef\delta="710E
\mathchardef\epsilon="710F
\mathchardef\zeta="7110
\mathchardef\eta="7111
\mathchardef\theta="7112
\mathchardef\iota="7113
\mathchardef\kappa="7114
\mathchardef\lambda="7115
\mathchardef\mu="7116
\mathchardef\nu="7117
\mathchardef\xi="7118
\mathchardef\pi="7119
\mathchardef\rho="711A
\mathchardef\sigma="711B
\mathchardef\tau="711C
\mathchardef\upsilon="711D
\mathchardef\phi="711E
\mathchardef\chi="711F
\mathchardef\psi="7120
\mathchardef\omega="7121
\mathchardef\varepsilon="7122
\mathchardef\vartheta="7123
\mathchardef\varpi="7124
\mathchardef\varrho="7125
\mathchardef\varsigma="7126
\mathchardef\varphi="7127
\mathchardef\imath="717B
\mathchardef\jmath="717C
\mathchardef\ell="7160
\mathchardef\wp="717D
\mathchardef\partial="7140
\mathchardef\flat="715B
\mathchardef\natural="715C
\mathchardef\sharp="715D


\def\err@badsizechange{%
	\immediate\write16{--> Size change not allowed in math mode, ignored}}

\baselinestretch=1000
\tenpoint

\catcode`\@=12					
\catcode`\@=11
\expandafter\ifx\csname @iasmacros\endcsname\relax
	\global\let\@iasmacros=\par
\else	\immediate\write16{}
	\immediate\write16{Warning:}
	\immediate\write16{You have tried to input iasmacros more than once.}
	\immediate\write16{}
	\endinput
\fi
\catcode`\@=12


\def\rmb{\seventeenrm}

\def\singlespace{\baselineskip=\normalbaselineskip}
\def\halfspace{\baselineskip=1.5\normalbaselineskip}
\def\doublespace{\baselineskip=2\normalbaselineskip}


\def\AB{\bigskip\parindent=40pt
        \centerline{\bf ABSTRACT}\medskip\halfspace\narrower}
\def\AE{\bigskip\nonarrower\doublespace}
\def\nonarrower{\advance\leftskip by-\parindent
	\advance\rightskip by-\parindent}


\def\boxit#1{\vbox{\hrule\hbox{\vrule\kern3pt
	\vbox{\kern3pt#1\kern3pt}\kern3pt\vrule}\hrule}}

\def\hence{\leavevmode\hbox{\bf .\raise5.5pt\hbox{.}.} }

\def\dalemb#1#2{{\vbox{\hrule height.#2pt
	\hbox{\vrule width.#2pt height#1pt \kern#1pt \vrule width.#2pt}
	\hrule height.#2pt}}}
\def\gtorder{\mathrel{\raise.3ex\hbox{$>$}\mkern-14mu
             \lower0.6ex\hbox{$\sim$}}}
\def\ltorder{\mathrel{\raise.3ex\hbox{$<$}\mkern-14mu
             \lower0.6ex\hbox{$\sim$}}}

\newdimen\fullhsize
\newbox\leftcolumn
\def\twoup{\hoffset=-.5in \voffset=-.25in
  \hsize=4.75in \fullhsize=10in \vsize=6.9in
  \def\fullline{\hbox to\fullhsize}
  \let\lr=L
  \output={\if L\lr
        \global\setbox\leftcolumn=\columnbox\global\let\lr=R \advancepageno
      \else \doubleformat \global\let\lr=L\fi
    \ifnum\outputpenalty>-20000 \else\dosupereject\fi}
  \def\doubleformat{\shipout\vbox{
    \fullline{\box\leftcolumn\hfil\columnbox}\advancepageno}}
  \def\columnbox{\leftline{\vbox{\makeheadline\pagebody\makefootline}}}
  \tolerance=1000 }

\twelvepoint
\doublespace
\overfullrule=0pt
{\nopagenumbers{
\rightline{IASSNS-HEP-99/83}
\rightline{~~~September, 1999}
\bigskip\bigskip
\centerline{\rmb Structure and Properties of Hughston's} 
\centerline{\rmb  Stochastic Extension of the Schr\"odinger Equation }
\medskip
\centerline{\it Stephen L. Adler}
\centerline{\bf Institute for Advanced Study}
\centerline{\bf Princeton, NJ 08540}
\medskip
\centerline{\it Lawrence P. Horwitz \footnote{*}
{\rm On leave from School of Physics and Astronomy, 
Raymond and Beverly Sackler Faculty 
of Exact Sciences, Tel Aviv University, Ramat Aviv, Israel, and 
Department of Physics, Bar Ilan University, Ramat Gan, Israel.}}
\centerline{\bf Institute for Advanced Study}
\centerline{\bf Princeton, NJ 08540}
\centerline{ }
\centerline{ }
\bigskip\bigskip
\leftline{\it Send correspondence to:}
\medskip
{\singlespace\leftline{Stephen L. Adler}
\leftline{Institute for Advanced Study}
\leftline{Olden Lane, Princeton, NJ 08540}
\leftline{Phone 609-734-8051; FAX 609-924-8399; email adler@ias.edu}}
\bigskip\bigskip
}}
\vfill\eject
\pageno=2
\AB
Hughston has recently proposed a stochastic extension of the Schr\"odinger 
equation, expressed as a stochastic differential equation on projective 
Hilbert space.  We derive new projective Hilbert space identities, which 
we use to give a general proof that Hughston's equation leads to 
state vector collapse to energy eigenstates, 
with collapse probabilities given by the quantum mechanical probabilities 
computed from the initial state.  We discuss the relation of Hughston's 
equation to earlier work on norm-preserving stochastic equations, and 
show that Hughston's equation can be written as a manifestly unitary 
stochastic evolution equation for the pure state density matrix.  We 
discuss the behavior of systems constructed as direct products of independent 
subsystems, and briefly address the question of whether an energy-based  
approach, such as Hughston's, suffices to give an objective interpretation 
of the measurement process in quantum mechanics.  
\AE                                       
\bigskip\bigskip
\vfill\eject
\pageno=3
\leftline{\bf I.~~INTRODUCTION}
\bigskip
A substantial body of work [1] has addressed the problem of state vector 
collapse by proposing that the Schr\"odinger equation be modified to 
include a stochastic process, presumably arising from physics at a deeper 
level, that drives the collapse  process.  In particular, Gisin [2], 
Percival [3], and Ghirardi, Pearle, and Rimini [4] have constructed 
equations that preserve the norm of the state vector, which in the 
approximation that the usual Schr\"odinger Hamiltonian dynamics is 
neglected are shown [4] to lead to state vector collapse with the 
correct quantum mechanical probabilities.  An alternative approach 
to constructing a stochastic extension of the Schr\"odinger equation has 
been pursued by Hughston [5], based on the proposal of a 
number of authors [6] to rewrite the Schr\"odinger 
equation as an equivalent dynamics on projective Hilbert space, i.e., on 
the space of rays, a formulation in which the imposition of a state vector 
normalization condition is not needed.  Within this framework, Hughston [5] 
has proposed a simple stochastic extension of the Schr\"odinger equation, 
constructed solely from the Hamiltonian function, and has shown that his 
equation leads to state vector reduction to an energy eigenstate, with   
energy conservation in the mean throughout the reduction process.    
In the simplest spin-1/2 case, Hughston exhibits an explicit solution 
that shows that his equation leads to collapse with the correct quantum 
mechanical probabilities, but the issue of collapse probabilities in the 
general case has remained open.   

Our purpose in this paper is to further investigate the structure 
and properties of Hughston's equation, proceeding from new identities in   
projective Hilbert space derived in Sec.~II.  
A principal result will be the proof in Sec.~III (using the martingale or 
``gambler's ruin'' argument pioneered by Pearle [7]) that 
in the generic case, with no approximations, Hughston's equation leads 
to state vector collapse to energy eigenstates with the 
correct quantum mechanical probabilities.  The relation of Hughston's 
equation to earlier work on norm-preserving equations is discussed in 
Sec.~IV, and the density matrix form of Hughston's equation, which gives 
a manifestly unitary stochastic evolution on pure states, is given in 
Sec.~V.  In Sec.~VI we examine the stochastic evolution of an initial 
state that is constructed as the product of independent subsystem states.  
Finally, in Sec.~VII we discuss whether an energy-based approach 
to stochastic evolution (as opposed to approaches [8] based on spontaneous 
localization) suffices to give a satisfactory objective description of 
the evolution of a state during the quantum mechanical measurement process.

\bigskip
\leftline{\bf II.~~PROJECTIVE HILBERT SPACE AND SOME IDENTITIES}
\bigskip

We begin by explaining the basic elements of projective Hilbert space 
needed to understand Hughston's 
equation, working in an $n+1$ dimensional Hilbert space.  We denote the 
general state vector in this space by $| z \rangle$, with $z$ a shorthand 
for the complex projections $z^0,z^1,...,z^n$ of the state vector on an 
arbitrary fixed basis.  Letting  $F$ be an arbitrary Hermitian operator, and 
using the summation convention that repeated indices are summed over their 
range, we define 
$$(F) \equiv { \langle z | F | z \rangle   \over \langle z |z \rangle } 
= { \overline z^{\alpha} F_{\alpha \beta} z^{\beta} \over  
\overline z^{\gamma} z^{\gamma} }~~~,
\eqno(1a)$$
so that $(F)$ is the expectation of the operator $F$ 
in the state $|z\rangle$, 
independent of the ray representative and normalization 
chosen for this state.  
Note that in this notation $(F^2)$ and $(F)^2$ are not the same; their 
difference is in fact the variance $[\Delta F]^2$, 
$$[\Delta F]^2 = (F^2)-(F)^2~~~.\eqno(1b)$$
We shall use two other parameterizations for the state $|z\rangle$ in what 
follows. Since $(F)$ is homogeneous of degree zero in both 
$z^{\alpha}$ and $\overline z^{\alpha}$, let us define new 
complex coordinates $t^j$ by  
$$t^j=z^j/z^0,~~ \overline t^j=\overline z^j 
/ \overline z^0~,~~~j=1,...,n, ~~~
\eqno(2)$$
which are well-defined over all states for which $z^0 \neq 0$ [9].  
Next, it is convenient to 
split each  of the complex numbers $t^j$ into its real and imaginary 
part $t^j_R,~t^j_I$, and to introduce a $2n$ component real vector 
$x^a,~a=1,...,2n$ defined by $x^1=t^1_R,~x^2=t^1_I,~x^3=t^2_R,~
x^4=t^2_I,...,x^{2n-1}=t^n_R,~x^{2n}=t^n_I$.   Clearly, specifying 
the projective coordinates $t^j$ or $x^a$ uniquely determines the 
unit ray containing the unnormalized state $|z\rangle$, while leaving 
the normalization and ray representative of the state $|z\rangle$ 
unspecified.   

As discussed in Refs. [6], projective Hilbert space is also a Riemannian  
space with respect to the Fubini-Study metric $g_{\alpha \beta}$, defined 
by the line element 
$$ds^2= g_{\alpha \beta} d\overline z^{\alpha} dz^{\beta}
\equiv 4\left( 1- { | \langle z | z+dz \rangle |^2 \over \langle z |z \rangle 
\langle z+dz | z+dz \rangle } \right) ~~~.\eqno(3a)$$ 
Abbreviating $\overline z^{\gamma} z^{\gamma} \equiv  \overline z \cdot z$, 
a simple calculation gives 
$$g_{\alpha \beta}=4(\delta_{\alpha \beta} \overline z \cdot z
-z^{\alpha} \overline z^{\beta})/(\overline z \cdot z)^2
=4 {\partial \over \partial \overline z^{\alpha} }
{\partial \over \partial z^{\beta} } \log \overline z \cdot z~~~.
\eqno(3b)$$
Because of the homogeneity conditions $\overline z^{\alpha} g_{\alpha \beta} 
=z^{\beta} g_{\alpha \beta}=0$, the metric $g_{\alpha \beta}$ is not 
invertible, but if we hold the coordinates $\overline z^0,~z^0$ fixed in  
the variation contained in  Eq.~(3a) and go over to the 
projective coordinates $t^j$, we can rewrite the line element of Eq.~(3a) 
as 
$$ds^2=g_{jk}d\overline t^j dt^k~~~,\eqno(4a)$$ 
with the invertible metric [9] 
$$g_{jk}={4[(1+\overline t^{\ell} t^{\ell}) \delta_{jk} - t^j \overline t^k ]
\over (1+\overline t^m t^m)^2 }~~~,\eqno(4b)$$                                 
                           
with inverse 
$$g^{jk}={1 \over 4} (1+\overline t^m t^m) (\delta_{jk} + t^j \overline t^k)
~~~.\eqno(4c)$$
Reexpressing the complex projective coordinates $t^j$ in terms of the 
real coordinates $x^a$, the line element can be rewritten as 
$$\eqalign{ 
ds^2=&g_{ab}dx^adx^b~~~,\cr
g_{ab}=&{4[(1+x^dx^d)\delta_{ab}-(x^ax^b+\omega_{ac}x^c\omega_{bd}x^d)] 
\over (1+x^e x^e)^2}~~~,\cr
g^{ab}=&{1\over 4} (1+x^e x^e)(\delta_{ab}+
x^ax^b+\omega_{ac}x^c\omega_{bd}x^d)~~~.\cr 
}\eqno(4d)$$
Here $\omega_{ab}$ is a numerical tensor whose only nonvanishing elements are  

$\omega_{a=2j-1 ~b=2j}=1$ and $\omega_{a=2j~b=2j-1}=-1$
for $j=1,...,n$.  As discussed 
by Hughston, one can define a complex structure $J_a^{~b}$ over the entire 
projective Hilbert space for which $J_a^{~c}J_b^{~d}g_{cd}=g_{ab},$   
$J_a^{~b}J_b^{~c}=-\delta_a^c$, 
such that $\Omega_{ab}=g_{bc} J_a^{~c}$ and 
$\Omega^{ab}=g^{ac}J_c^{~b}$ are antisymmetric tensors.  At $x=0$, the metric 
and complex structure take the values 
$$\eqalign{
g_{ab}=&4 \delta_{ab}~,~~g^{ab}={1 \over 4} \delta_{ab}~~~,\cr
J_a^{~b}=&\omega_{ab}~,~~\Omega_{ab}=4\omega_{ab}~,
~~\Omega^{ab}={1\over 4}\omega_{ab}~~~.\cr
}\eqno(5)$$

Returning to Eq.~(1a), we shall now derive some identities that are 
central to what follows.  Differentiating Eq.~(1a) with respect to 
$\overline z^{\alpha}$, with respect to $z^{\beta}$, and with respect to both 
$\overline z^{\alpha}$ and $z^{\beta}$, we get 
$$\eqalign{
\langle z | z \rangle {\partial (F) \over \partial \overline z^{\alpha}}
=&F_{\alpha \beta} z^{\beta} - (F) z^{\alpha}~~~,\cr
\langle z | z \rangle {\partial (F) \over \partial  z^{\beta}}=&
\overline z^{\alpha} F_{\alpha \beta}  - (F) \overline z^{\beta}~~~,\cr
\langle z | z \rangle^2 {\partial^2 (F) \over \partial \overline z^{\alpha} 
\partial z^{\beta} }=&
\langle z |z \rangle [F_{\alpha \beta}-\delta_{\alpha \beta} (F) ]
+2z^{\alpha} \overline z^{\beta} (F) - \overline z^{\gamma} F_{\gamma \beta} 
z^{\alpha}-\overline z^{\beta} F_{\alpha \gamma} z^{\gamma}~~~.\cr
}\eqno(6a)$$
Writing similar expressions for a second operator expectation $(G)$, 
contracting in various combinations with the relations of Eq.~(6a), and 
using the homogeneity conditions 
$$
\overline z^{\alpha} {\partial (F) \over \partial \overline z^{\alpha} }
=z^{\beta} {\partial (F) \over \partial z^{\beta} }
=\overline z^{\alpha} {\partial^2 (F) \over \partial \overline z^{\alpha} 
\partial z^{\beta}}
=z^{\beta} {\partial^2 (F) \over  \partial \overline z^{\alpha} 
\partial z^{\beta} } =0~~~\eqno(6b)$$
to eliminate derivatives with respect to $\overline z^0,~z^0$, we get 
the following identities,
$$\eqalign{
-i(FG-GF)=-i \langle z| z \rangle \left( 
{\partial (F) \over \partial z^{\alpha}} 
{\partial (G) \over \partial \overline z^{\alpha}} - 
{\partial (G) \over \partial z^{\alpha}} 
{\partial (F) \over \partial \overline z^{\alpha}} \right) 
=&2\Omega^{ab} \nabla_a (F) \nabla_b (G)~~~,\cr
(FG+GF)-2(F)(G)= \langle z| z \rangle \left( 
{\partial (F) \over \partial z^{\alpha}} 
{\partial (G) \over \partial \overline z^{\alpha}} + 
{\partial (G) \over \partial z^{\alpha}} 
{\partial (F) \over \partial \overline z^{\alpha}} \right) 
=&2g^{ab} \nabla_a (F) \nabla_b (G)~~~,\cr
(FGF)-(F^2)(G)-(F)(FG+GF)+2(F)^2(G)=\langle z | z \rangle ^2 
{\partial (F) \over \partial z^{\alpha}}
{\partial^2 G \over \partial \overline z^{\alpha} \partial z^{\beta}}
{\partial (F) \over \partial \overline z^{\beta}}
=&2\nabla^a (F) \nabla^b (F) \nabla_a \nabla_b (G),\cr
}\eqno(7a)$$
with $\nabla_a$ the covariant derivative constructed using the Fubini-Study 
metric affine connection.  It is not necessary to use 
the detailed form of this affine connection 
to verify the right hand equalities in these identities, because since $(G)$ 
is a Riemannian scalar, $\nabla_a \nabla_b (G)$$ =\nabla_a \partial_b (G)$, 
and since projective Hilbert space is a homogeneous manifold, it suffices 
to verify the identities at the single point $x=0$, where the affine 
connection vanishes and thus 
$\nabla_a \nabla_b (G)=\partial_a \partial_b (G)$.
Using Eqs.~(7a) and the chain rule we also find 
$$-\nabla_a [(F^2)-(F)^2] \nabla^a (G)=
-{1\over 2} (F^2 G+G F^2) +(F^2)(G) + (F) (FG+GF) -2(F)^2(G)~~~.
\eqno(7b)$$
When combined with the final identity in Eq.~(7a) this gives 
$$\eqalign{
D\equiv&\nabla^a(F) \nabla^b(F) \nabla_a \nabla_b (G)
-{1\over 2} \nabla_a [(F^2)-(F)^2] \nabla^a (G)\cr
=&{1\over 4}(2FGF-F^2G-GF^2)\cr
=&-{1\over 4}([F,[F,G]])~~~,\cr
}\eqno(7c)$$
with $[~,~]$ denoting the commutator, from which we see 
that $D$ vanishes when the operators $F$ and $G$ commute.  

An alternative derivation of Eq.~(7c) proceeds from the fact, noted by 
Hughston, that (for $F$ self-adjoint) 
$$\xi^a_F\equiv \Omega^{ab} \nabla_b(F)~~~\eqno(8a) $$ 
is a 
Killing vector obeying 
$$\nabla_c\xi_F^a+\nabla^a \xi_{cF}=0~~~.\eqno(8b)$$  
Using the identity $(F^2)-(F)^2=\nabla_b(F) \nabla^b(F)$, which is the 
$F=G$ case of the middle equality of Eq.~(7a), we rewrite $D$ of 
Eq.~(7c) as 
$$D=\nabla^a(F) \nabla^b(F) \nabla_a\nabla_b(G)
-\nabla^b(F)\nabla^a(G)\nabla_a \nabla_b(F)~~~.\eqno(9a)$$
This can be rewritten, using the identity 
$\Omega^{ab}\Omega_{cb}=\delta^a_c$, the antisymmetry of $\Omega$,  
the fact that $\Omega$ commutes with the covariant derivatives, 
and the Killing vector definition of Eq.~(8a), as 
$$\eqalign{
D=&\Omega^{ac} \xi_{cF} \Omega^{be} \xi_{eF} \nabla_a \nabla_b (G)
-\Omega^{bc} \xi_{cF} \Omega^{ae} \xi_{eG} \nabla_a \nabla_b (F)\cr
=&-\xi_{cF} \xi_{eF} \Omega^{ac} \nabla_a \xi^e_G
+\xi_{cF}\xi_{eG}\Omega^{ae}\nabla_a\xi^c_F~~~.
}\eqno(9b)$$
We now use the Killing vector identity of Eq.~(8b) on the final factor  
in each term, giving 
$$D=\xi_{cF} \xi_{eF} \Omega^{ac} \nabla^e \xi_{aG}
-\xi_{cF}\xi_{eG}\Omega^{ae}\nabla^c\xi_{aF}~~~.\eqno(9c)$$
Exchanging the labels $e$ and $c$ in the first term, and exchanging the 
labels $a$ and $e$ in the second term, we get 
$$\eqalign{
D=&\xi_{cF} \xi_{eF} \Omega^{ae} \nabla^c \xi_{aG}
+\xi_{cF}\xi_{aG}\Omega^{ae}\nabla^c\xi_{eF} \cr
=&\xi_{cF} \nabla^c[\Omega^{ae}\xi_{eF}\xi_{aG}]~~~.\cr  
}\eqno(10a)$$
Substituting the Killing vector definition of Eq.~(8a), this becomes   
$$\eqalign{
D=&\Omega_{cb}\nabla^b(F)\nabla^c[\Omega^{ae}
\Omega_{ef}\nabla^f(F)\Omega_{ag}\nabla^g(G)]\cr  
=&\Omega_{cb}\nabla^b(F)\nabla^c[\Omega_{gf}\nabla^g(G)\nabla^f(F)]\cr
=&-{1\over 4}([F,[F,G]])~~~,\cr
}\eqno(10b)$$
where to get the final line we have twice used the first identity in 
Eq.~(7a).   This completes our geometric derivation of Eq.~(7c). 
\bigskip  
\vfill\break
\leftline{\bf III.~~HUGHSTON'S EQUATION AND STATE VECTOR} 
\leftline{\bf ~~~~~~COLLAPSE PROBABILITIES}
\bigskip

Let us now turn to Hughston's stochastic differential equation, which 
reads  
$$dx^a=[2 \Omega^{ab}\nabla_b(H)-{1\over 4}\sigma^2 \nabla^aV]dt
+\sigma\nabla^a(H) dW_t~~~,\eqno(11a)$$
with $W_t$ a Brownian motion or Wiener process, with $\sigma$ a parameter 
governing the strength of the stochastic terms, with $H$ the Hamiltonian 
operator and $(H)$ its expectation, and with  $V$ the 
variance of the Hamiltonian, 
$$V=[\Delta H]^2=(H^2)-(H)^2~~~.\eqno(11b)$$
When the parameter $\sigma$ is zero, Eq.~(11a) is just [6] the transcription 
of the Schr\"odinger equation to projective 
Hilbert space.  For the time evolution of a 
general function $G[x]$, we get by 
Taylor expanding $G[x+dx]$ and using the It\^o stochastic calculus rules [10]
$$[dW_t]^2=dt~,~~[dt]^2=dtdW_t=0~~~,\eqno(12a)$$
the corresponding stochastic differential equation
$$dG[x]=\mu dt + \sigma \nabla_aG[x]\nabla^a(H) dW_t~~~,\eqno(12b)$$
with the drift term $\mu$ given by 
$$\mu=2 \Omega^{ab} \nabla_aG[x]\nabla_b(H)-{1\over 4}
\sigma^2\nabla^aV\nabla_a
G[x]+{1\over 2}\sigma^2 \nabla^a(H)\nabla^b(H)\nabla_a\nabla_bG[x]~~~.
\eqno(12c)$$
Hughston shows that with the $\sigma^2$ part of the drift term 
chosen as in Eq.~(11a), the drift term $\mu$ in Eq.~(12b) vanishes for the 
special case $G[x]=(H)$, guaranteeing conservation 
of the expectation of the energy with respect to the stochastic evolution 
of Eq.~(11a).  But referring to Eq. (7c) and the first identity in Eq.~(7a),  
we see that in fact 
a much stronger result is also true, namely that $\mu$ vanishes [and thus
the stochastic process of Eq.~(12b) is a martingale] whenever 
$G[x]=(G)$, with $G$ any operator that commutes with the Hamiltonian $H$.   

Let us now make two applications of this fact.  First, taking $G[x]=V=
(H^2)-(H)^2$, we see that the contribution from $(H^2)$ to $\mu$ 
vanishes, so the drift term comes entirely from $-(H)^2$.  
Substituted this into $\mu$ gives $-2(H)$ times the drift term produced by 
$(H)$, which is again zero, plus an extra term 
$$-\sigma^2 \nabla^a(H)\nabla ^b(H)\nabla_a(H)\nabla_b(H)
=-\sigma^2V^2~~~,\eqno(13a)$$
where we have used the relation $V=\nabla_a(H)\nabla^a(H)$ which follows 
from the $F=G=H$ case of the middle identity of Eq.~(7a).  Thus the 
variance $V$ of the Hamiltonian satisfies the stochastic differential 
equation, derived by Hughston by a more complicated method, 
$$dV=-\sigma^2 V^2 dt + \sigma \nabla_aV\nabla^a(H) dW_t~~~.\eqno(13b)$$  
This implies that the expectation $E[V]$ with respect to the stochastic 
process obeys 
$$E[V(t)]=E[V(0)]-\sigma^2 \int_0^t ds E[V(s)^2]~~~,\eqno(13c)$$
which using the inequality $0\leq E[\{V-E[V]\}^2]=E[V^2]-E[V]^2$ 
gives the inequality 
$$E[V(t)] \leq E[V(0)] -\sigma^2 \int_0^t ds E[V(s)]^2~~~.\eqno(13d)$$
Since $V$ is necessarily positive, Eq.~(13d) implies that $E[V(\infty)]=0$, 
and again using positivity of $V$ this implies that $V(s)$ 
vanishes as $s \to \infty$, apart from a set of outcomes 
of probability measure zero.  Thus, as concluded by 
Hughston, the stochastic term in his equation drives the system, as $t \to 
\infty$, to an energy eigenstate.  

As our second application of the vanishing of the drift term $\mu$ for 
expectations of operators that commute with $H$, let us consider the 
projectors $\Pi_e\equiv |e\rangle \langle e| $ on a complete set of 
energy eigenstates $|e \rangle$.  By definition, these projectors all 
commute with H, and so the drift term $\mu$ vanishes in the stochastic 
differential equation for $G[x]=(\Pi_e)$, and consequently the expectations   
$E[(\Pi_e)]$ are time independent; additionally, by completeness of   
the states $|e\rangle$, we have $\sum_e (\Pi_e)=1$.  
But these are just the conditions for   
Pearle's [7] gambler's ruin argument to apply.  At time zero, 
$E[(\Pi_e)]=(\Pi_e)\equiv p_e$
is the absolute value squared of the quantum mechanical amplitude  
to find the initial state in energy eigenstate $|e \rangle$.  At $t=\infty$, 
the system always evolves to an energy eigenstate, with the eigenstate  
$|f\rangle $ occurring with some probability $P_f$.  The expectation 
$E[(\Pi_e)]$, evaluated at infinite time, is then  
$$E[(\Pi_e)]=1 \times P_e + \sum_{f \neq e} 0 \times P_f = P_e~~~;\eqno(14)$$
hence $p_e=P_e$ for each  $e$ and the state collapses into energy eigenstates 
at $t=\infty$ with probabilities given by the usual quantum mechanical 
rule applied to the initial wave function.  

This conclusion clearly 
generalizes to the stochastic equation 
$$dx^a=[2 \Omega^{ab}\nabla_b(H)-{1\over 4}\sigma^2\sum_j \nabla^aV_j]dt
+\sigma\sum_j\nabla^a(H_j) dW_t^j~~~,\eqno(15a)$$
with the $H_j$ a set of mutually commuting self-adjoint 
operators that commute 
with $H$, with $V_j=(H_j^2)-(H_j)^2$, and with the $dW_t^j$ independent 
Wiener processes obeying $dW_t^jdW_t^k=\delta^{jk}dt$.  
Following the same method used in obtaining Eq.~(13b), and defining  
$C_{kj}=\nabla_a(H_k)\nabla^a(H_j)
=(H_kH_j)-(H_k)(H_j)$, one finds
$$dV_k=-\sigma^2 \sum_j C_{kj}^2 dt + \sigma \nabla_aV_k
\sum_j \nabla^a(H_j) dW_t^j~~~,\eqno(15b)$$  
and therefore 
$$E[V_k(t)]=E[V_k(0)]-\sigma^2 \int_0^t ds \sum_j E[C_{kj}(s)^2]~~~.
\eqno(15c)$$
Since $E[C_{kj}^2] \geq E[C_{kj}]^2$, we have as before 
$$E[V_k(t)] \leq E[V_k(0)]-\sigma^2 \int_0^t ds \sum_j 
E[C_{kj}(s)]^2~~~,\eqno(15d)$$
which implies that each $E[C_{kj}(s)]$ approaches zero as 
$s \to \infty$.  
Hence for each $k,j$ we have at large times 
$$E[(H_kH_j)]-E[(H_k)(H_j)] \to 0  ~~~,\eqno(15e)$$  
and so there is an effective quantum decorrelation of commuting observables.  
Moreover, for $k=j$ Eq.~(15e) implies that at large times $E[V_k] \to 0$,   
which since $V_k$ is nonnegative implies that $V_k$ approaches zero apart 
from a set of outcomes of probability measure zero,   
and so the state evolves to a simultaneous eigenstate of all the commuting 
observables entering the process of Eq.~(15a).
\bigskip
\leftline{\bf IV.~~RELATION OF HUGHSTON'S EQUATION TO OTHER}  
\leftline{\bf ~~~~ STOCHASTIC NORM-PRESERVING EQUATIONS}
\bigskip

Let us now specialize Eqs.~(12b) and (12c) to the case in which $G[x]$ is  
simply the expectation $(G)$ of an operator $G$.  Then by substituting  
Eqs.~(7c) and the second equality in Eq.~(7a), we find 
$$d(G)=\mu dt + \kappa dW_t~~~,\eqno(16a)$$
with 
$$\kappa={1\over 2}\sigma[(\{G,H\})-2(G)(H)]
={1\over 2}\sigma(\{G,H-(H)\})~~~,\eqno(16b)$$ 
and with
$$\mu=(-i[G,H])-{1\over 8}\sigma^2([H,[H,G]])~~~,\eqno(16c)$$
where we have used $\{~,~\}$ to denote the anticommutator.  

Let us now compare this with the evolution of $(G)$ implied by the 
stochastic state vector evolution 
$$d|z\rangle=[\alpha dt + \beta dW_t] |z\rangle ~~~,\eqno(17a)$$
with 
$$\eqalign{
\alpha=&-iH-{1\over 8}\sigma^2[A-(A)]^2~~~,\cr
\beta=&{1\over 2} \sigma [A-(A)]~~~,\cr
}\eqno(17b)$$
where $A$ is a general self-adjoint operator and $(A)$ is defined, as in 
Eq.~(1a), by $(A)=\langle z|A|z\rangle / \langle z|z \rangle$.  
For the evolution of 
$\langle z |G|z \rangle$, we find by the It\^o rules, 
$$\eqalign{
d\langle z |G|z \rangle =& \langle z | 
[\alpha^{\dagger} G + G \alpha+\beta^{\dagger} G \beta]dt 
+ [\beta^{\dagger}G+G \beta] dW_t |z\rangle\cr
=&\langle z|  -i[G,H]dt -{1\over 8}\sigma^2
[\{G,[A-(A)]^2\}-2[A-(A)]G[A-(A)]]dt\cr
&~~~~~~~~~+{1\over 2}\sigma \{G,[A-(A)]\} dW_t |z\rangle\cr  
=&\langle z| -i[G,H]dt -{1\over 8}\sigma^2[A,[A,G]]dt
+{1\over 2}\sigma \{G,[A-(A)]\} dW_t |z\rangle~~~.\cr  
}\eqno(18a)$$
When $G=1$, the right hand side of Eq.~(18a) vanishes, since the commutator 
terms vanish trivially and $\langle z|A-(A)|z \rangle=\langle z|A|z \rangle 
-\langle z|z \rangle (A)=0$.  Therefore the state vector evolution of 
Eqs.~(17a, b) is norm preserving, and so it is consistent to choose the   
normalization $\langle z|z \rangle =1$ in conjunction with this evolution.  
For general $G$  we then have $d\langle z |G|z \rangle=d(G)$, and 
so Eq.~(18a) gives   
an expression for $d(G)$, which we see is identical to Eqs.~(16a) - 
(16c) when the operator $A$ is taken as the Hamiltonian $H$.  In particular, 
when $A=G=H$ we learn from Eq.~(18a) that  
$d(H)=\sigma V dW_t$, in agreement with 
Eq.~(12b), because
$$\langle z| \{H,H-(H)\}|z\rangle = 2[(H^2)-(H)^2]=2V~~~,\eqno(18b)$$ 
and so the convergence argument of Eqs.~(13a)-(13d) follows directly 
from Eq.~(18a).  Apart from 
minor changes in notation, the norm preserving evolution of Eqs.~(17a) 
and (17b) is the one given by Gisin [2], Percival [3], and 
Ghirardi, Pearle, and Rimini [4], and so we see that this evolution is 
equivalent [11] to the state vector evolution in projective 
Hilbert space given by Hughston's equation.  

The evolution of Eq.~(17a) 
can be generalized [as was done for the Hughston equation in Eq.~(15)] 
to read
$$\eqalign{
d|z\rangle=&[\alpha dt +\sum_j \beta_j dW_t^j] |z\rangle~~~,\cr 
\alpha=&-iH-{1\over 8}\sigma^2\sum_j[A_j-(A_j)]^2~~~,\cr
\beta_j=&{1\over 2} \sigma [A_j-(A_j)]~~~,\cr
}\eqno(19)$$
with the $A_j$ any set of mutually commuting operators.  When the $A_j$ 
do not all commute with the Hamiltonian $H$, it is necessary to make the 
approximation of neglecting the Hamiltonian evolution (the $-iH$ term 
in $\alpha$) in proving that Eq.~(19) implies state vector reduction to 
the mutual eigenstates of the $A_j$ with probabilities given by the usual 
quantum mechanical rule.  Such a proof, very similar to the one given 
for Hughston's equation in Sec.~III above, has been given by 
Ghirardi, Pearle, and Rimini [4].  In order to carry through the proof 
with no approximations, it is necessary to assume that the $A_j$ are 
operators in the mutually commuting set $H_j$ that all commute with $H$, 
as was done in Sec. III.
\bigskip
\leftline{\bf V.~~DENSITY MATRIX EVOLUTION}

Let us now define the pure state density matrix $\rho$ by 
$$\rho={ |z\rangle \langle z| \over \langle z|z \rangle },~~~\eqno(20a)$$
in terms of which $(G)$ is given by 
$$(G)={\rm Tr} \rho G~~~.\eqno(20b)$$
Since $G$ is a fixed operator, Eq.~(18a) for $d(G)$ can be rewritten as  
$$\eqalign{
{\rm Tr} G d\rho =&{\rm Tr} \rho\left[ -i[G,H]dt 
-{1\over 8}\sigma^2[A,[A,G]]dt
+{1\over 2}\sigma \{G,[A-(A)]\} dW_t \right]\cr
=&{\rm Tr} G \left[ -i[H,\rho]dt   -{1\over 8}\sigma^2[A,[A,\rho]]dt
+{1\over 2}\sigma \{\rho,[A-(A)]\} dW_t \right]~~~,\cr
}\eqno(20c)$$
where in the final line we have cyclically permuted terms under the trace.  
Since Eq.~(20c) holds for arbitrary self-adjoint operators $G$, it implies 
that the density matrix obeys the stochastic differential equation (each 
term of which is self-adjoint) 
$$d\rho= -i[H,\rho]dt   -{1\over 8}\sigma^2[A,[A,\rho]]dt
+{1\over 2}\sigma \{\rho,[A-(A)]\} dW_t ~~~.\eqno(21a)$$
This equation can be written in an alternative form by observing that 
since $\rho$ is a pure state density matrix obeying $\rho^2=\rho$, we 
have $\rho(A)=\rho  {\rm Tr}\rho A=\rho A \rho$.  These facts imply that  
$$\eqalign{
\{\rho,[A-(A)]\}=& \rho A+A \rho -2 \rho(A)\cr
=&\rho^2 A + A \rho^2 - 2 \rho A \rho = [\rho,[\rho,A]~~~,\cr
}\eqno(21b)$$
and so we can rewrite Eq.~(21a) as  
$$d\rho= -i[H,\rho]dt   -{1\over 8}\sigma^2[A,[A,\rho]]dt         
+{1\over 2}\sigma [\rho,[\rho,A]] dW_t~~~. \eqno(21c)$$

Equations (21a) and (21c) have the following properties for 
general $A$:\hfill\break
\item{(i)}  Since ${\rm Tr} d\rho=0$, the condition ${\rm Tr} \rho=1$
is preserved by the time evolution.  
\item{(ii)}  After some algebra using the It\^o rules, one finds 
that $\rho^2=\rho$ implies that 
$$\{\rho,d\rho\}+[d\rho]^2=d\rho~~~,\eqno(21d)$$ 
which can be rewritten as $[\rho+d\rho]^2=\rho+d\rho$. Hence the evolution  
of Eqs.~(21a, c) is consistent with the pure state condition.  This is  
required by the fact that Eqs.~(21a, c) may be derived as consequences 
of Eqs.~(17a, b), which are a norm preserving pure state evolution.    
The condition of Eq.~(21d) determines the coefficient of the $\sigma^2...dt$ 
drift term in terms of the coefficient of the $\sigma...dW_t$ stochastic
term, and so the ratio of these two coefficients in Eq.~(21c) cannot be 
treated as an additional adjustable parameter.  
\item{(iii)}  Since $d\rho=d\rho^{\dagger}$, the self-adjointness of $\rho$ 
is preserved by the time evolution.  
\item{(iv)}  Time reversal invariance is violated by the stochastic terms, 
since when $dt$ and $i$ are reversed in sign,  the term $ -i[H,\rho]dt$ is  
invariant, but the term $-{1\over 8}\sigma^2[A,[A,\rho]]dt$ reverses sign. 
\item{(v)}  When we take the stochastic expectation of Eq.~(21c), the 
$dW_t$ term drops out, and we get 
$${dE[\rho] \over dt}=-i[H,E[\rho]] -{1\over 8}\sigma^2[A,[A,E[\rho]]]~~~, 
\eqno(22a)$$
which as pointed out by Percival [3] and Ghirardi, Pearle, and Rimini [4] 
is a quantum dynamical semigroup evolution of the completely 
positive Lindblad [12] form.  The stochastic expectation $E[\rho]$  is 
what is usually termed the density matrix; it starts off at $t=0$ as a 
pure state density matrix but then evolves, through the stochastic 
process, into a mixed state density matrix.  
\item{(vi)}  The conditions for Eq.~(22a) 
to admit stationary solutions $E[\rho]_S$ with $dE[\rho]_S/dt=0$ are very 
stringent, since when the left hand side of Eq.~(22a) is zero, multiplying 
by $E[\rho]_S$ and taking the trace gives 
$$0=-i {\rm Tr} E[\rho]_S [H,E[\rho]_S] 
-{1\over 8} \sigma^2 {\rm Tr} E[\rho]_S [A,[A,E[\rho]_S]] ~~~.\eqno(22b)$$
Using cyclic permutation under the trace, the first term on the right hand 
side vanishes, while the second term becomes 
$${1\over 8} \sigma^2 {\rm Tr} [A,E[\rho]_S]^2~~~,\eqno(22c)$$
which can only vanish when $[A,E[\rho]_S]=0$.  Substituting this 
equation back into Eq.~(22a) then further implies that $[H,E[\rho]_S]=0$.  
When $A$ and $H$ commute, these conditions can be satisfied with  
$E[\rho]_S$ a general function of $H$ [see the further discussion of 
this case in (xiii) below] , but when $A$ and $H$ do not 
commute,  one can have situations in which either $E[\rho]_S$  must 
be a multiple of the unit operator, which trivially commutes with both 
$H$ and $A$, or else there are no stationary 
solutions and $E[\rho]$ diverges at large times.  The latter case is  
found in spontaneous localization models, as discussed for 
example in Section III.B.3 of Ref. [4].  
\item{(vii)}  The evolution 
implied by  Eq.~(22a) leads [13] to a monotonic increase of the 
von Neumann entropy,  a result that can be demonstrated directly from  
Eq.(22a) as follows.   Letting 
$$S=-{\rm Tr} E[\rho] \log E[\rho]~~~\eqno(22d)$$
be the von Neumann  (or information) entropy, we find by 
substituting Eq.~(22a) and   
using cyclical permutation of factors under the trace, that 
$$\eqalign{
{dS \over dt}=& -{\rm Tr} {dE[\rho] \over dt} [1+\log E[\rho] ] \cr
=&{1\over 8} \sigma^2 {\rm Tr}[\log E[\rho],A][A,E[\rho]]~~~.\cr
}\eqno(22e)$$
Substituting  complete sets of eigenstates $|n\rangle,~|m\rangle$ 
of the nonnegative density 
matrix $E[\rho]$, this becomes 
$$\eqalign{
{dS \over dt}=&{1 \over 8} \sigma^2 \sum_n\sum_m 
[\log E[\rho]_n - \log E[\rho]_m]A_{nm} A_{mn} [E[\rho]_n-E[\rho]_m]  \cr
=&{1 \over 8} \sigma^2 \sum_n\sum_m 
|A_{nm}|^2 [\log E[\rho]_n - \log E[\rho]_m] [E[\rho]_n-E[\rho]_m] \geq 0
~~~.\cr
}\eqno(22f)$$
\item{(viii)}  Since Eqs.~(21a) and (21c) are nonlinear in $\rho$, the 
Schr\"odinger dynamics described by them cannot be represented as an 
equivalent Heisenberg or dual dynamics on the operator $G$.  On the other 
hand, Eq.~(22a) is linear in $\rho$, and so as noted by Lindblad, the 
Schr\"odinger dynamics for $E[\rho]$ can be represented as a dual Heisenberg 
dynamics for $E[G]$, given by 
$${dE[G] \over dt}=i[H,E[G]]-{1\over 8}\sigma^2[A,[A,E[G]]]~~~. 
\eqno(22g)$$
\item{(ix)}  The evolution of Eq.~(21c) can be written (after some 
algebra, and again using $\rho^2=\rho$) in the manifestly 
unitary form 
$$\rho+d\rho=U\rho U^{\dagger}~,~~~U=e^{dK}~~~,\eqno(23a)$$
with the infinitesimal anti-self-adjoint generator $dK$ given by 
$$dK=\left[ -iH-{1\over 8} \sigma^2[A^2-2A\rho A,\rho]\right] dt
-{1\over 2}\sigma[\rho,A]dW_t~~~.\eqno(23b)$$
Equations (21c) and (23b)  thus give the stochastic unitary extension of the 
Lindblad evolution of Eq.~(22a) [14].

Specializing to the Hughston case $A=H$, Eq.~(21c) (which uses the 
pure state condition $\rho^2=\rho$) becomes 
$$d\rho= -i[H,\rho]dt   -{1\over 8}\sigma^2[H,[H,\rho]]dt         
+{1\over 2}\sigma [\rho,[\rho,H]] dW_t~~~,\eqno(24a)$$
while Eq.~(22a) becomes 
$${dE[\rho] \over dt}=-i[H,E[\rho]] -{1\over 8}\sigma^2[H,[H,E[\rho]]]~~~, 
\eqno(24b)$$
and the following further properties are evident:\hfill\break
\item{(x)}  When $\rho=\Pi_e$, the projector on an energy eigenstate, 
then since all commutators in Eq.~(24a) vanish we have $d\rho=0$. 
\item{(xi)} For $G$ commuting with $H$, $E[d(G)]={\rm Tr }G dE[\rho]=0$, 
since by cyclic permutation inside the trace each term arising from 
substituting Eq.~(24b) into the expectation of Eq.~(20b) 
can be rearranged 
to have a factor $[G,H]$.  
\item{(xii)} For $V=[\Delta H]^2={\rm Tr} \rho H^2 - [{\rm Tr}\rho H]^2$, 
use of Eq.~(24a) and the It\^o calculus imply that $E[dV/dt]=-E[R^2]$, with 
$$R={1\over 2} \sigma {\rm Tr} [\rho,[\rho,H]] H =\sigma V~~~.\eqno(24c)$$
\item{(xiii)} Items (x) through (xii) are the ingredients used in Sec.~III 
to prove state vector collapse to energy eigenstates $|e\rangle$ 
with the correct quantum 
mechanical probabilities $p_e$. Hence at large times, as noted by 
Hughston, $E[\rho] \to \sum_e p_e \Pi_e$, which explicitly exhibits the  
role of $E[\rho]$ as the density matrix that evolves, under the stochastic 
process, from a pure to a mixed state form.  The fact that $E[\rho]$ for 
Hughston's equation approaches a stationary limit at large times 
is in accord with the general stationarity discussion given in (vi) above.  
\vfill\eject
\bigskip 
\leftline{\bf VI.~~BEHAVIOR OF SYSTEMS CONSTRUCTED} 
\leftline{\bf ~~~~~FROM INDEPENDENT SUBSYSTEMS}
\bigskip

Let us next examine the structure of Hughston's 
equation for a Hilbert space constructed as the direct product of  
independent subsystem Hilbert spaces, so that initially at time $t=0$ 
the state vector is 
$$|z\rangle = \prod_{\ell} |z_{\ell} \rangle~~~. \eqno(25a)$$   
We assume the Hamiltonian 
$$H=\sum_{\ell} H_{\ell}~~~,\eqno(25b)$$
with $H_{\ell}$ acting as the unit operator on the 
states $|z_{k}\rangle ~,~~
k \neq \ell$.  Then a simple calculation shows that the expectation 
of the Hamiltonian $(H)$ and its variance $V$  are both 
additive over the subsystem Hilbert spaces, 
$$\eqalign{
(H)=\sum_{\ell} (H_{\ell})_{\ell}~~~,\cr
V=\sum_{\ell} V_{\ell} =\sum_{\ell}[ (H_{\ell}^2)_{\ell} 
-(H_{\ell})_{\ell}^2]~~~,\cr
}\eqno(25c)$$
with $(F_{\ell})_{\ell}$ the expectation of the operator $F_{\ell}$ 
formed according to Eq.~(1a) with respect 
to the subsystem wave function $|z_{\ell}\rangle$.  In addition, 
the Fubini-Study line element is also additive over the subsystem Hilbert 
spaces, since 
$$\eqalign{
1-ds^2/4=& {| \langle z | z+dz \rangle |^2 \over \langle z |z \rangle 
\langle z+dz | z+dz \rangle } =\prod_{\ell}  
{ | \langle z_{\ell} | z_{\ell}+dz_{\ell} \rangle |^2 \over 
\langle z_{\ell} |z_{\ell} \rangle \langle z_{\ell}+dz_{\ell} 
| z_{\ell}+dz_{\ell} \rangle }\cr 
=&\prod_{\ell}[1-ds_{\ell}^2/4]=1-[\sum_{\ell} ds_{\ell}^2]/4 +{\rm O}(ds^4)
~~~.\cr}\eqno(26)$$
[An alternative way to see this is to use the identity 
$\log \overline z \cdot z =\log \prod_{\ell} \overline z_{\ell} 
\cdot z_{\ell} =$ $ \sum_{\ell} \log \overline z_{\ell}\cdot z_{\ell}$ 
in Eq.~(3b), along with a change of 
variable from $z$ to the $z_{\ell}$'s.]
As a result of Eq.~(26), the metric $g^{ab}$ and complex  
structure $\Omega^{ab}$ block diagonalize over the independent subsystem 
subspaces. Equations (25a)-(25c) then imply that 
Hughston's stochastic extension  
of the Schr\"odinger equation given in Eq.~(11a) 
separates into similar equations for the subsystems, that do not refer 
to one another's $x^a$ coordinates, but are correlated only through the  
common Wiener process $dW_t$ that appears in all of them.  These correlations 
result in the entanglement of the states $|z_{\ell}\rangle$, 
so that the product 
form of Eq.~(25a) is not maintained for times $t>0$, but subsystems 
$|z_{\ell}\rangle$ already in energy eigenstates 
remain unentangled for all time,  
since the coefficient of $dW_t$ vanishes in their stochastic evolution 
equations.  

These same conclusions follow from the density matrix form of Hughston's 
equation given in Eq.~(24a), in which the entanglements arising from the 
action of the same Wiener process on all subsystems are already evident, 
because the density matrix depends quadratically on the normalized state 
vector.  Considering for simplicity the case of two independent subsystems, 
substituting the $t=0$ form 
$$\rho = \rho_1 \rho_2 ~~~\eqno(27a)$$   
into Eq.~(24a), with $H=H_1+H_2$, we get
$$d\rho=d\rho_1 \rho_2+\rho_1 d\rho_2 
-{1 \over 4}\sigma^2 [H_1,\rho_1][H_2,\rho_2]dt  
~~~,\eqno(27b)$$
with $d\rho_1$ the evolution predicted by Eq.~(24a) within subsystem 1, 
$$d\rho_1= -i[H_1,\rho_1]dt   -{1\over 8}\sigma^2[H_1,[H_1,\rho]]dt         
+{1\over 2}\sigma [\rho_1,[\rho_1,H_1]] dW_t~~~,\eqno(27c)$$
and similarly for $d\rho_2$.  The entangling term proportional to 
$ [H_1,\rho_1][H_2,\rho_2]dt$ comes from the $[dW_t]^2$ contribution from   
the state vector evolution equation to the density matrix equation; it 
is in general nonzero, but vanishes when either $[H_1,\rho_1]=0$ or 
$[H_2,\rho_2]=0$, that is, when either of the two subsystems is in an 
energy eigenstate.  When more than two subsystems are present, the 
entangling term coupling $\rho_L$ to $\rho_{\ell}~,~~~\ell \neq L$ is 
more complicated in structure, but still has a factor  $[H_L,\rho_L]$
and so vanishes when the subsystem $L$ is in an energy eigenstate.  Thus 
the endpoints of the stochastic evolution under Hughston's equation, which 
are the energy eigenstates, can persist indefinitely as unentangled 
independent subsystems in a larger system.  

This conclusion does not extend to the more general 
evolution of Eq.~(21c), in which the stochastic process 
is driven by an operator $A$ differing from the Hamiltonian, with $A$ 
taken to be additive over subsystems.  The reason is that there is 
now a competition between the stochastic terms, which are constructed  
from double commutators with an innermost commutator $[A,\rho]$, and  
the Schr\"odinger evolution term, which involves the commutator $[H,\rho]$; 
the stochastic terms tend to drive the system to $A$ eigenstates, while 
the Schr\"odinger term  coherently mixes $A$ eigenstates, leading to   
evolution away from $A$ eigenstates.  Thus, a subsystem cannot 
remain indefinitely in an $A$ eigenstate, and 
as a result does not persist indefinitely 
as an unentangled independent subsystem in a larger system.  [These 
statements are in accord with the conclusions reached in the stationarity 
discussion of (vi) in Sec. V.]

\bigskip
\leftline{\bf VII.  DOES AN ENERGY-BASED EQUATION SUFFICE?}
\bigskip

In the preceding sections we have seen how Hughston's equation fits 
into the general framework of stochastic modifications of the 
Schr\"odinger equation that have been studied in the past.  Its 
distinguishing feature is that the general operator $A$ of Eqs.~(17a, b) and 
(21a, c) driving the                                                         
stochastic terms is chosen, in Hughston's case, to be the Hamiltonian $H$.  
This choice confers the advantage that the proof of reduction of the 
state vector to $A$ eigenstates (i.e., in Hughston's case, to energy  
eigenstates) with the correct quantum mechanical probabilities becomes 
exact, since it is not necessary to neglect the Hamiltonian evolution term.  
Moreover, since for Hughston's equation 
the stochastic expectation of the Hamiltonian operator 
$E[H]$ is conserved in time, and since convergence to $H$ eigenstates 
preserves the quantum mechanical predictions, 
any statistical test of energy conservation 
performed on the endpoint of the stochastic process will agree with the 
quantum mechanical prediction.  To justify these advantages, we must 
now address the issue of whether an energy-based stochastic equation is 
sufficient to give an objective interpretation of state 
vector reduction [15].  

First, we must deal with the objection that in most measurements, the 
quantum attribute being measured is not an energy; for example, in a 
Stern-Gerlach experiment, it is typically the $z$ component of a spin.  
However, to perform a measurement, it is always necessary to couple 
the quantum attribute being measured to the apparatus through an interaction 
energy term $H_I$, in such a way that the macroscopic state of the 
apparatus is ultimately determined by the quantum attribute being measured.  
Thus, in the first instance, what is being measured is an energy, even 
though after amplification to macroscopic scale this can be converted 
to other forms of indication, such as pointer displacements.  So from 
the point of view of the variety of quantum attributes that can be 
measured, Hughston's equation appears to be as viable as localizing 
approaches [8] in which $A$ is chosen as an operator that produces 
spatial localization.  

We must next deal with the issue of whether an energy-based approach can 
prevent the occurrence of macroscopic quantum superpositions.  For example, 
take a macroscopic object and displace it a macroscopic distance; the 
two states have the same energy, and so in Hughston's approach such 
superpositions would appear to be allowed, whereas in localizing approaches 
they are strongly forbidden.  However, this objection neglects the 
interactions of the macroscopic object with its environment, of the same  
type that are important in studies of decoherence.  When such effects are 
taken into account, macroscopic displacement of a macroscopic object 
results in an energy shift $\Delta E$, reflecting the altered environment,  
which is sufficient, from the 
point of view of Hughston's equation, to lead to rapid state vector 
reduction to one displaced alternative or the other.  To study this 
quantitatively, let us consider the following two environmental effects: 
(i) thermal energy fluctuations, and (ii) the surface adsorption of 
surrounding molecules.  Hughston proposes, as have other authors [16], that 
the  parameter governing the stochastic terms is of order 
$\sigma \sim M_{\rm Planck}^{-1/2}$ 
in microscopic units with $\hbar =c=1$, which he shows leads to   
state vector reduction in a time $t_R$ given by 
$$t_R \sim \left( { 2.8 {\rm MeV} \over \Delta E } \right)^2 {\rm sec}~~~.
\eqno (28)$$
Hence to get a reduction time of order, say, $10^{-6}$ seconds, one needs 
a  $\Delta E \sim 3 {\rm GeV} \sim 3~{\rm nucleon~masses}$.  

Considering first the effect of thermal fluctuations, let us consider 
a macroscopic object with $N\sim 10^{23}$ nucleon masses, 
so that $\Delta E \sim N^{1\over 2} kT \sim 8 {\rm GeV}$ at room temperature
($300^{\circ}$ Kelvin) and $\Delta E \sim .08 {\rm Gev}$ at the $3^{\circ}$  
temperature of the cosmic microwave background.  For such an object, 
thermal energy driven state vector reduction will occur in $10^{-7}$ seconds 
at room temperature and in $10^{-3}$ seconds at the temperature of the 
microwave background.  Examining next the effect of adsorbed molecules, 
consider an object with a surface area of $1~{\rm cm}^2$ at room temperature 
in an extreme vacuum of $10^{-14} {\rm Torr}$  (less [17] than the nighttime 
pressure at the surface of the moon.)   Then the flux of 
molecules bombarding its surface is [17] $4 \times 10^6$ per second, so 
assuming a high probability for the molecules to stick, 
a $\Delta E$ of $3 {\rm GeV}$ is attained in of order $10^{-6}$ seconds,    
permitting a $10^{-6}$ second state vector reduction time driven 
by the change in 
energy produced by surface adsorption.  One can scale to 
other sizes of macroscopic object from these examples, but they suffice 
to show that in the normal range of laboratory operating conditions for 
measuring apparatus, environmental interactions produce a large enough  
spread of energy values to give rapid state vector reduction through an 
energy driven stochastic equation.  

>From a formal point of view, it is instructive to cast the above discussion 
of environmental effects in terms of the analysis of the measurement 
process given by Zurek [18], starting from Eq.~(24b) for the evolution 
of the stochastic 
expectation of the density matrix. 
Zurek assumes that the total Hamiltonian $H$ describes the system $\cal S$ 
being measured, the apparatus $\cal  A$ doing the measuring, and the 
environment $\cal E$.  Thus, he writes the Hamiltonian as a sum of 6 terms, 
$$H=H_{\cal S}+H_{\cal A}+H_{\cal E}+H_{\cal SA}+H_{\cal AE}+H_{\cal SE}~~~,
\eqno(29)$$
with the first three terms giving the Hamiltonians of the system, apparatus, 
and environment in isolation from one another, and with the second three 
terms giving the corresponding interaction Hamiltonians.  Zurek assumes 
that the interaction $H_{\cal SE}$ between system and environment can be 
neglected, and that the interaction $H_{\cal SA}$ between system and 
apparatus acts only briefly while entanglement of the system and apparatus 
states is established, but is unimportant during the subsequent evolution 
of the density matrix that results in the actual measurement.  He also 
makes the simplifying assumption that the states which actually distinguish 
between quantities being measured have equal 
eigenvalues of the non-interaction part of the 
Hamiltonian $H_{\cal S}+H_{\cal A}+H_{\cal E}$, which    
implies that for the submatrix of $E[\rho]$ spanned by these states, the 
commutator $[H_{\cal S}+H_{\cal A}+H_{\cal E},E[\rho]]$ is zero, and so   
these commutator terms in Eq.~(24b) can be neglected.  With 
these simplifications, Eq.~(24b) becomes
$${dE[\rho] \over dt}=-i[H_{\cal AE},E[\rho]] 
-{1\over 8}\sigma^2[H_{\cal AE},[H_{\cal AE},E[\rho]]]~~~, 
\eqno(30a)$$
or when the non-Schr\"odinger term is omitted, as in Zurek's analysis, 
$${dE[\rho] \over dt}=-i[H_{\cal AE},E[\rho]] ~~~.\eqno(30b)$$
Zurek points out that the evolution of Eq.~(30b) introduces 
correlations between the apparatus and the environment, which select as 
the ``pointer basis'' of the apparatus, that registers the measurement,  
the eigenstates $|A_p\rangle$ of a ``pointer observable'' $\hat \Pi$ 
that commutes with $H_{\cal AE}$; in other words, the   
pointer basis projectors $\Pi_p=|A_p\rangle \langle A_p|$ must satisfy 
$$ [\Pi_p,H_{\cal AE}]=0~~~.\eqno(30c)$$
Returning to the full evolution equation of Eq.~(30a), 
with the non-Schr\"odinger terms included, we see that the 
argument of Sec.~III, 
when applied to this equation using Eq.~(30c), implies state vector 
collapse to the eigenstates of the Zurek pointer basis.    
Thus an energy-based stochastic reduction 
equation, when analyzed within the framework of Zurek's approximations, 
is consistent with, and adds further support to, 
the picture of the measurement process that 
Zurek proposes in [18].    

In addition to the issues just discussed, there are further  
questions that must be addressed  
in an energy-based approach, such as whether Hughston's estimated $\sigma$ 
gives sufficiently rapid (but also not too rapid) reduction of state 
vectors for all classes of experiments that have been carried out.  Answering 
this question is beyond the scope of the present paper, but is an 
important issue for future study.   Ultimately, the decision between an 
energy-based or localization-based approach (or yet some other choice of 
the operator $A$ driving the stochastic terms) may depend on which form 
of the modified Schr\"odinger equation can be derived as an  
approximation to relativistically invariant physics at a deeper level.  

To summarize, we have shown that Hughston's stochastic extension of the 
Schr\"odinger equation has properties that make it a viable physical model 
for state vector reduction.   This opens the challenge of seeing whether 
it can be derived as a phenomenological approximation to a fundamental 
pre-quantum dynamics, along the lines of existing work on open dynamical   
systems [19].  Specifically, we suggest that since 
Adler and Millard [20] have argued that quantum mechanics can emerge as 
the thermodynamics of an underlying non-commutative operator dynamics, 
and since the corrections to the thermodynamic approximation in this dynamics 
are driven by the trace of the energy operator 
multiplied by a coefficient parameter with dimensions of inverse mass, 
it may be possible to show that Hughston's stochastic process is the  
leading statistical fluctuation correction to this thermodynamics.

\bigskip

\centerline{\bf Acknowledgments}
This work was supported in part by the Department of Energy under
Grant \#DE--FG02--90ER40542.  One author (S.L.A.) wishes to thank J. 
Anandan for conversations introducing him to the Fubini-Study metric,    
F. Benatti for a conversation about evolutions of the Lindblad type, and 
G. C. Ghirardi and A. Bassi for emphasizing the relevance to our  
discussion of Ref.~[4] and for a stimulating discussion.   He also wishes 
to acknowledge the hospitality of the Aspen Center for Physics, where this 
manuscript was completed.  
The other author (L.P.H.) wishes 
to thank P. Leifer for many discussions on the properties of the complex 
projective manifold, and D. Moore for helpful conversations on this work.  
He is grateful to C. Piron of the University of Geneva, and the CERN Theory 
Division, for their hospitality during the final stages of this work.  
Helpful comments from referees are also acknowledged 
with appreciation.  

\vfill\eject
\centerline{\bf References}
\bigskip
\noindent
[1]  For a representative, but not exhaustive, survey of the earlier 
literature, see the papers of Di\'osi, Ghirardi et. al., Gisin, 
Pearle, and Percival cited by Hughston, Ref. [5] below.
\hfill\break
\medskip 
\noindent
[2]  N. Gisin, Helv. Phys. Acta {\bf 62}, 363 (1989).\hfill\break
\medskip
\noindent
[3]  I. C. Percival, Proc. R. Soc. Lond. A{\bf447}, 189 (1994).\hfill\break
\medskip
\noindent
[4]  G. C. Ghirardi, P. Pearle, and A. Rimini, Phys. Rev. A{\bf 42}, 
78 (1990).\hfill\break
\medskip
\noindent
[5]  L. P. Hughston, Proc. Roy. Soc. Lond. A {\bf 452}, 953 (1996).
\hfill\break
\medskip
\noindent
[6]  D. A. Page, Phys. Rev. A {\bf 36}, 3479 (1987); Y. Aharanov and 
J. Anandan, Phys. Rev. Lett. {\bf 58}, 1593 (1987); J. Anandan and Y. 
Aharanov, Phys. Rev. D {\bf 38}, 1863 (1988) 
and Phys. Rev. Lett. {\bf 65}, 1697 (1990); G. W. Gibbons, 
J. Geom. Phys. {\bf 8}, 147 (1992); L. P. Hughston, ``Geometric aspects 
of quantum mechanics'', in S. A. Huggett, ed., {\it Twistor theory}, 
Marcel Dekker, New York, 1995; A. Ashtekar and T. A. Schilling, preprint 
gr-qc/9706069.  For related work, see  
A. Heslot, Phys. Rev. D {\bf 31}, 1341 (1985) 
and S. Weinberg, Phys. Rev. Lett. {\bf 62}, 485 (1989) and 
Ann. Phys. (NY) {\bf 194}, 336 (1989).  \hfill\break
\medskip
\noindent
[7]  P. Pearle, Phys. Rev. D {\bf 13}, 857 (1976); Phys. Rev. D {\bf 29}, 
235 (1984); Phys. Rev. A {\bf 39}, 2277 (1989).\hfill\break  
\medskip
\noindent
[8]  See, e.g., Ref.~[4] above and G. C. Ghirardi, A. Rimini, and 
T. Weber, Phys. Rev. D {\bf 34}, 470 (1986).\hfill\break  
\medskip
\noindent
[9]  There are states which have some other element, say $z^k \neq 0$, 
and which overlap with those for which $z^0 \neq 0$. The function $(F)$,  
expressed in terms of the $t^j$'s, can then be extended by continuity 
to a function of a second set of variables $t^{j\prime} = z^j/z^k$, 
defined over the set $z^k \neq 0$.  With this process, one can extend  
the function $(F)$ to the covering projective space.
Similarly, in Eqs. (3a)-(4b), what we have called $z^0$ could 
be any $z^{\alpha}\neq 0$.  There is 
therefore a set of holomorphically overlapping patches, so that the 
metric of Eq.~(4b) is globally defined.  See, for example, S. Kobayashi 
and K. Nomizu, {\it Foundations of Differential Geometry}, Vol. II, p. 159, 
Wiley Interscience, New York, 1969. \hfill\break
\medskip
\noindent
[10]  For an excellent exposition of the It\^o calculus, see C. W. Gardiner, 
{\it Handbook of Stochastic Methods}, Springer-Verlag, Berlin, 1990, 
Chapt. 4.   \hfill\break
\medskip
\noindent
[11]  We wish to thank A. Bassi and G. C. Ghirardi for a conversation 
explaining this connection.  Our exposition closely follows theirs.\hfill  
\break
\medskip
\noindent
[12]  G. Lindblad, Commun. Math. Phys. {\bf 48}, 119 (1976); V. Gorini, 
A. Kossakowski, and E. C. G. Sudarshan, J. Math. Phys. {\bf 17}, 821 
(1976).  Abbreviating $\overline \rho \equiv E[\rho]$, the most general 
Lindblad-type evolution is 
$${d \overline \rho \over dt}= -i[H,\overline \rho]
+\sum_j[V_j\overline \rho V_j^{\dagger}- {1\over 2} V_j^{\dagger}V_j 
\overline \rho - {1\over 2} \overline \rho V_j^{\dagger}V_j]~~~,$$
which when $V_j^{\dagger}=V_j$ reduces to 
$${d \overline \rho \over dt}= -i[H,\overline \rho]
-{1\over 2}\sum_j[V_j,[V_j,\overline \rho]] ~~~,$$
corresponding to the structure of Eq.~(22a).  The positivity results 
of (vi) and (vii) below are special to the case of self-adjoint $V_j$, 
and do not extend to the general Lindblad-type evolution with 
$V_j^{\dagger} \neq V_j$. \hfill\break
\medskip
\noindent
[13]  For a general discussion, see G. Lindblad, {\it Non-Equilibrium 
Entropy and Irreversibility}, D. Reidel, Dordrecht, 1983, pp. 28-29; 
for a recent $2 \times 2$ matrix 
derivation and application to $K$-meson decays, see 
F. Benatti and R. Floreanini, in ``Quantum Probability'', 
Banach Center Publications, Vol. 43, Institute of Mathematics, 
Polish Academy of Sciences, Warsaw, 1998.\hfill\break  
\medskip
\noindent
[14]  We wish to thank F. Benatti for calling our attention to the 
book K. R. Parasarathy, 
{\it An Introduction to the Quantum Stochastic Calculus},  
Birkh\"auser Verlag, Basel, 1992, Chapt. III,
which discusses such extensions.\hfill\break
\medskip
\noindent
[15] For earlier discussions of energy-based reduction, see G. J. Milburn, 
Phys. Rev. A {\bf 44}, 5401 (1991); D. Bedford and D. Wang, Nuovo Cimento 
{\bf 26}B, 313 (1975) and Nuovo Cimento {\bf 37}B, 55 (1977).  \hfill\break
\medskip
\noindent
[16]  See L. P. Hughston, Ref. [5], Sec. 11 and earlier 
work of Di\'osi, Ghirardi 
et. al., and Penrose cited there; also D. I. Fivel, preprint 
quant-ph/9710042.\hfill\break  
\medskip
\noindent
[17]  P. A. Redhead, article on {\it Vacuum} in the Macmillan  
Encyclopedia of Physics, J. G. Rigden, ed., Simon \& Schuster Macmillan, 
New York, 1996, p. 1657.
\hfill\break
\medskip
\noindent
[18]  W. H. Zurek, Phys. Rev. D {\bf 24}, 1516 (1981).\hfill\break
\medskip
\noindent
[19]  See, e.g., H. Spohn, Rev. Mod. Phys. {\bf 53}, 569 (1980).\hfill\break
\medskip
\noindent
[20] S. L. Adler and A. C. Millard, Nucl. Phys. B {\bf 473}, 199 (1966); 
see also S. L. Adler and A. Kempf, J. Math. Phys. {\bf 39}, 5083 (1998).
\hfill\break
\bigskip
\noindent
\bye
\bigskip
\noindent
\bigskip
\noindent
\bigskip
\noindent
\bigskip
\noindent
\bigskip
\noindent
\bigskip
\noindent
\bigskip
\noindent
\bigskip
\noindent
\bigskip
\noindent
\bigskip
\noindent
\bigskip
\noindent
\bigskip
\noindent
\bigskip
\noindent
\bigskip
\noindent
\vfill
\eject
\bigskip
\bye

Return-Path: adler@sns.ias.edu 
Received: from thunder.sns.ias.edu (thunder [198.138.243.12])
	by blackhole.sns.ias.edu (8.8.5/8.8.5) with ESMTP id LAA11024
	for <val@sns.ias.edu>; Wed, 8 Sep 1999 11:27:30 -0400 (EDT)
Received: from thunder (adler@localhost)
	by thunder.sns.ias.edu (8.8.5/8.8.5) with ESMTP id LAA22917
	for <val>; Wed, 8 Sep 1999 11:27:28 -0400 (EDT)
Message-Id: <199909081527.LAA22917@thunder.sns.ias.edu>
X-Authentication-Warning: thunder.sns.ias.edu: adler owned process doing -bs
X-Mailer: exmh version 1.6.7 5/3/96
To: val@sns.ias.edu
Mime-Version: 1.0
Date: Wed, 08 Sep 1999 11:27:28 -0400
From: Stephen Adler <adler@sns.ias.edu>
Content-Type: text/plain; charset=us-ascii
Content-Length: 66573

Dear Val -  Please post this on the quant-phy  Bulletin Board, with 
cross listing to hep-th.
Thanks, 

Steve
______________________________________________________________________

\catcode`\@=11					



\font\fiverm=cmr5				
\font\fivemi=cmmi5				
\font\fivesy=cmsy5				
\font\fivebf=cmbx5				

\skewchar\fivemi='177
\skewchar\fivesy='60


\font\sixrm=cmr6				
\font\sixi=cmmi6				
\font\sixsy=cmsy6				
\font\sixbf=cmbx6				

\skewchar\sixi='177
\skewchar\sixsy='60


\font\sevenrm=cmr7				
\font\seveni=cmmi7				
\font\sevensy=cmsy7				
\font\sevenit=cmti7				
\font\sevenbf=cmbx7				

\skewchar\seveni='177
\skewchar\sevensy='60


\font\eightrm=cmr8				
\font\eighti=cmmi8				
\font\eightsy=cmsy8				
\font\eightit=cmti8				
\font\eightbf=cmbx8				

\skewchar\eighti='177
\skewchar\eightsy='60


\font\ninei=cmmi9
\font\ninesy=cmsy9

\skewchar\ninei='177
\skewchar\ninesy='60


\font\tenrm=cmr10				
\font\teni=cmmi10				
\font\tensy=cmsy10				
\font\tenex=cmex10				
\font\tenit=cmti10				
\font\tensl=cmsl10				
\font\tenbf=cmbx10				
\font\tentt=cmtt10				
\font\tenss=cmss10				
\font\tensc=cmcsc10				
\font\tenbi=cmmib10				

\skewchar\teni='177
\skewchar\tenbi='177
\skewchar\tensy='60

\def\tenpoint{\ifmmode\err@badsizechange\else
	\textfont0=\tenrm \scriptfont0=\sevenrm \scriptscriptfont0=\fiverm
	\textfont1=\teni  \scriptfont1=\seveni  \scriptscriptfont1=\fivemi
	\textfont2=\tensy \scriptfont2=\sevensy \scriptscriptfont2=\fivesy
	\textfont3=\tenex \scriptfont3=\tenex   \scriptscriptfont3=\tenex
	\textfont4=\tenit \scriptfont4=\sevenit \scriptscriptfont4=\sevenit
	\textfont5=\tensl
	\textfont6=\tenbf \scriptfont6=\sevenbf \scriptscriptfont6=\fivebf
	\textfont7=\tentt
	\textfont8=\tenbi \scriptfont8=\seveni  \scriptscriptfont8=\fivemi
	\def\rm{\tenrm\fam=0 }%
	\def\it{\tenit\fam=4 }%
	\def\sl{\tensl\fam=5 }%
	\def\bf{\tenbf\fam=6 }%
	\def\tt{\tentt\fam=7 }%
	\def\ss{\tenss}%
	\def\sc{\tensc}%
	\def\bmit{\fam=8 }%
	\rm\setparameters\setbaselines\fi}


\font\twelverm=cmr12				
\font\twelvei=cmmi12				
\font\twelvesy=cmsy10	scaled\magstep1		
\font\twelveex=cmex10	scaled\magstep1		
\font\twelveit=cmti12				
\font\twelvesl=cmsl12				
\font\twelvebf=cmbx12				
\font\twelvett=cmtt12				
\font\twelvess=cmss12				
\font\twelvesc=cmcsc10	scaled\magstep1		
\font\twelvebi=cmmib10	scaled\magstep1		

\skewchar\twelvei='177
\skewchar\twelvebi='177
\skewchar\twelvesy='60

\def\twelvepoint{\ifmmode\err@badsizechange\else
	\textfont0=\twelverm \scriptfont0=\eightrm \scriptscriptfont0=\sixrm
	\textfont1=\twelvei  \scriptfont1=\eighti  \scriptscriptfont1=\sixi
	\textfont2=\twelvesy \scriptfont2=\eightsy \scriptscriptfont2=\sixsy
	\textfont3=\twelveex \scriptfont3=\tenex   \scriptscriptfont3=\tenex
	\textfont4=\twelveit \scriptfont4=\eightit \scriptscriptfont4=\sevenit
	\textfont5=\twelvesl
	\textfont6=\twelvebf \scriptfont6=\eightbf \scriptscriptfont6=\sixbf
	\textfont7=\twelvett
	\textfont8=\twelvebi \scriptfont8=\eighti  \scriptscriptfont8=\sixi
	\def\rm{\twelverm\fam=0 }%
	\def\it{\twelveit\fam=4 }%
	\def\sl{\twelvesl\fam=5 }%
	\def\bf{\twelvebf\fam=6 }%
	\def\tt{\twelvett\fam=7 }%
	\def\ss{\twelvess}%
	\def\sc{\twelvesc}%
	\def\bmit{\fam=8 }%
	\rm\setparameters\setbaselines\fi}


\font\fourteenrm=cmr12	scaled\magstep1		
\font\fourteeni=cmmi12	scaled\magstep1		
\font\fourteensy=cmsy10	scaled\magstep2		
\font\fourteenex=cmex10	scaled\magstep2		
\font\fourteenit=cmti12	scaled\magstep1		
\font\fourteensl=cmsl12	scaled\magstep1		
\font\fourteenbf=cmbx12	scaled\magstep1		
\font\fourteentt=cmtt12	scaled\magstep1		
\font\fourteenss=cmss12	scaled\magstep1		
\font\fourteensc=cmcsc10 scaled\magstep2	
\font\fourteenbi=cmmib10 scaled\magstep2	

\skewchar\fourteeni='177
\skewchar\fourteenbi='177
\skewchar\fourteensy='60

\def\fourteenpoint{\ifmmode\err@badsizechange\else
	\textfont0=\fourteenrm \scriptfont0=\tenrm \scriptscriptfont0=\sevenrm
	\textfont1=\fourteeni  \scriptfont1=\teni  \scriptscriptfont1=\seveni
	\textfont2=\fourteensy \scriptfont2=\tensy \scriptscriptfont2=\sevensy
	\textfont3=\fourteenex \scriptfont3=\tenex \scriptscriptfont3=\tenex
	\textfont4=\fourteenit \scriptfont4=\tenit \scriptscriptfont4=\sevenit
	\textfont5=\fourteensl
	\textfont6=\fourteenbf \scriptfont6=\tenbf \scriptscriptfont6=\sevenbf
	\textfont7=\fourteentt
	\textfont8=\fourteenbi \scriptfont8=\tenbi \scriptscriptfont8=\seveni
	\def\rm{\fourteenrm\fam=0 }%
	\def\it{\fourteenit\fam=4 }%
	\def\sl{\fourteensl\fam=5 }%
	\def\bf{\fourteenbf\fam=6 }%
	\def\tt{\fourteentt\fam=7}%
	\def\ss{\fourteenss}%
	\def\sc{\fourteensc}%
	\def\bmit{\fam=8 }%
	\rm\setparameters\setbaselines\fi}


\font\seventeenrm=cmr10 scaled\magstep3		


\newdimen\rp@
\newcount\@basestretchnum
\newskip\@baseskip
\newskip\headskip
\newskip\footskip


\def\setparameters{\rp@=.1em
	\headskip=24\rp@
	\footskip=\headskip
	\delimitershortfall=5\rp@
	\nulldelimiterspace=1.2\rp@
	\scriptspace=0.5\rp@
	\abovedisplayskip=10\rp@ plus3\rp@ minus5\rp@
	\belowdisplayskip=10\rp@ plus3\rp@ minus5\rp@
	\abovedisplayshortskip=5\rp@ plus2\rp@ minus4\rp@
	\belowdisplayshortskip=10\rp@ plus3\rp@ minus5\rp@
	\normallineskip=\rp@
	\lineskip=\normallineskip
	\normallineskiplimit=0pt
	\lineskiplimit=\normallineskiplimit
	\jot=3\rp@
	\setbox0=\hbox{\the\textfont3 B}\p@renwd=\wd0
	\skip\footins=12\rp@ plus3\rp@ minus3\rp@
	\skip\topins=0pt plus0pt minus0pt}


\def\setbaselines{\maxdepth=4\rp@\baselinestretch=\@basestretchnum}


\def\baselinestretch{\afterassignment\@basestretch\@basestretchnum}
\def\@basestretch{%
	\@baseskip=12\rp@ \divide\@baseskip by1000
	\normalbaselineskip=\@basestretchnum\@baseskip
	\baselineskip=\normalbaselineskip
	\bigskipamount=\the\baselineskip
		plus.25\baselineskip minus.25\baselineskip
	\medskipamount=.5\baselineskip
		plus.125\baselineskip minus.125\baselineskip
	\smallskipamount=.25\baselineskip
		plus.0625\baselineskip minus.0625\baselineskip
	\setbox\strutbox=\hbox{\vrule height.708\baselineskip
		depth.292\baselineskip width0pt }}



\def\makeheadline{\vbox to0pt{\baselinestretch=1000
	\vskip-\headskip \vskip1.5pt
	\line{\vbox to\ht\strutbox{}\the\headline}\vss}\nointerlineskip}

\def\makefootline{\baselineskip=\footskip\line{\the\footline}}

\def\big#1{{\hbox{$\left#1\vbox to8.5\rp@ {}\right.\n@space$}}}
\def\Big#1{{\hbox{$\left#1\vbox to11.5\rp@ {}\right.\n@space$}}}
\def\bigg#1{{\hbox{$\left#1\vbox to14.5\rp@ {}\right.\n@space$}}}
\def\Bigg#1{{\hbox{$\left#1\vbox to17.5\rp@ {}\right.\n@space$}}}


\mathchardef\alpha="710B
\mathchardef\beta="710C
\mathchardef\gamma="710D
\mathchardef\delta="710E
\mathchardef\epsilon="710F
\mathchardef\zeta="7110
\mathchardef\eta="7111
\mathchardef\theta="7112
\mathchardef\iota="7113
\mathchardef\kappa="7114
\mathchardef\lambda="7115
\mathchardef\mu="7116
\mathchardef\nu="7117
\mathchardef\xi="7118
\mathchardef\pi="7119
\mathchardef\rho="711A
\mathchardef\sigma="711B
\mathchardef\tau="711C
\mathchardef\upsilon="711D
\mathchardef\phi="711E
\mathchardef\chi="711F
\mathchardef\psi="7120
\mathchardef\omega="7121
\mathchardef\varepsilon="7122
\mathchardef\vartheta="7123
\mathchardef\varpi="7124
\mathchardef\varrho="7125
\mathchardef\varsigma="7126
\mathchardef\varphi="7127
\mathchardef\imath="717B
\mathchardef\jmath="717C
\mathchardef\ell="7160
\mathchardef\wp="717D
\mathchardef\partial="7140
\mathchardef\flat="715B
\mathchardef\natural="715C
\mathchardef\sharp="715D


\def\err@badsizechange{%
	\immediate\write16{--> Size change not allowed in math mode, ignored}}

\baselinestretch=1000
\tenpoint

\catcode`\@=12					
\catcode`\@=11
\expandafter\ifx\csname @iasmacros\endcsname\relax
	\global\let\@iasmacros=\par
\else	\immediate\write16{}
	\immediate\write16{Warning:}
	\immediate\write16{You have tried to input iasmacros more than once.}
	\immediate\write16{}
	\endinput
\fi
\catcode`\@=12


\def\rmb{\seventeenrm}

\def\singlespace{\baselineskip=\normalbaselineskip}
\def\halfspace{\baselineskip=1.5\normalbaselineskip}
\def\doublespace{\baselineskip=2\normalbaselineskip}


\def\AB{\bigskip\parindent=40pt
        \centerline{\bf ABSTRACT}\medskip\halfspace\narrower}
\def\AE{\bigskip\nonarrower\doublespace}
\def\nonarrower{\advance\leftskip by-\parindent
	\advance\rightskip by-\parindent}


\def\boxit#1{\vbox{\hrule\hbox{\vrule\kern3pt
	\vbox{\kern3pt#1\kern3pt}\kern3pt\vrule}\hrule}}

\def\hence{\leavevmode\hbox{\bf .\raise5.5pt\hbox{.}.} }

\def\dalemb#1#2{{\vbox{\hrule height.#2pt
	\hbox{\vrule width.#2pt height#1pt \kern#1pt \vrule width.#2pt}
	\hrule height.#2pt}}}
\def\gtorder{\mathrel{\raise.3ex\hbox{$>$}\mkern-14mu
             \lower0.6ex\hbox{$\sim$}}}
\def\ltorder{\mathrel{\raise.3ex\hbox{$<$}\mkern-14mu
             \lower0.6ex\hbox{$\sim$}}}

\newdimen\fullhsize
\newbox\leftcolumn
\def\twoup{\hoffset=-.5in \voffset=-.25in
  \hsize=4.75in \fullhsize=10in \vsize=6.9in
  \def\fullline{\hbox to\fullhsize}
  \let\lr=L
  \output={\if L\lr
        \global\setbox\leftcolumn=\columnbox\global\let\lr=R \advancepageno
      \else \doubleformat \global\let\lr=L\fi
    \ifnum\outputpenalty>-20000 \else\dosupereject\fi}
  \def\doubleformat{\shipout\vbox{
    \fullline{\box\leftcolumn\hfil\columnbox}\advancepageno}}
  \def\columnbox{\leftline{\vbox{\makeheadline\pagebody\makefootline}}}
  \tolerance=1000 }

\input iasmacros
\twelvepoint
\doublespace
\overfullrule=0pt
{\nopagenumbers{
\rightline{IASSNS-HEP-99/36}
\rightline{~~~September, 1999}
\bigskip\bigskip
\centerline{\rmb Structure and Properties of Hughston's} 
\centerline{\rmb  Stochastic Extension of the Schr\"odinger Equation }
\medskip
\centerline{\it Stephen L. Adler}
\centerline{\bf Institute for Advanced Study}
\centerline{\bf Princeton, NJ 08540}
\medskip
\centerline{\it Lawrence P. Horwitz \footnote{*}
{\rm On leave from School of Physics and Astronomy, 
Raymond and Beverly Sackler Faculty 
of Exact Sciences, Tel Aviv University, Ramat Aviv, Israel, and 
Department of Physics, Bar Ilan University, Ramat Gan, Israel.}}
\centerline{\bf Institute for Advanced Study}
\centerline{\bf Princeton, NJ 08540}
\centerline{ }
\centerline{ }
\bigskip\bigskip
\leftline{\it Send correspondence to:}
\medskip
{\singlespace\leftline{Stephen L. Adler}
\leftline{Institute for Advanced Study}
\leftline{Olden Lane, Princeton, NJ 08540}
\leftline{Phone 609-734-8051; FAX 609-924-8399; email adler@ias.edu}}
\bigskip\bigskip
}}
\vfill\eject
\pageno=2
\AB
Hughston has recently proposed a stochastic extension of the Schr\"odinger 
equation, expressed as a stochastic differential equation on projective 
Hilbert space.  We derive new projective Hilbert space identities, which 
we use to give a general proof that Hughston's equation leads to 
state vector collapse to energy eigenstates, 
with collapse probabilities given by the quantum mechanical probabilities 
computed from the initial state.  We discuss the relation of Hughston's 
equation to earlier work on norm-preserving stochastic equations, and 
show that Hughston's equation can be written as a manifestly unitary 
stochastic evolution equation for the pure state density matrix.  We 
discuss the behavior of systems constructed as direct products of independent 
subsystems, and briefly address the question of whether an energy-based  
approach, such as Hughston's, suffices to give an objective interpretation 
of the measurement process in quantum mechanics.  
\AE                                       
\bigskip\bigskip
\vfill\eject
\pageno=3
\leftline{\bf I.~~INTRODUCTION}
\bigskip
A substantial body of work [1] has addressed the problem of state vector 
collapse by proposing that the Schr\"odinger equation be modified to 
include a stochastic process, presumably arising from physics at a deeper 
level, that drives the collapse  process.  In particular, Gisin [2], 
Percival [3], and Ghirardi, Pearle, and Rimini [4] have constructed 
equations that preserve the norm of the state vector, which in the 
approximation that the usual Schr\"odinger Hamiltonian dynamics is 
neglected are shown [4] to lead to state vector collapse with the 
correct quantum mechanical probabilities.  An alternative approach 
to constructing a stochastic extension of the Schr\"odinger equation has 
been pursued by Hughston [5], based on the proposal of a 
number of authors [6] to rewrite the Schr\"odinger 
equation as an equivalent dynamics on projective Hilbert space, i.e., on 
the space of rays, a formulation in which the imposition of a state vector 
normalization condition is not needed.  Within this framework, Hughston [5] 
has proposed a simple stochastic extension of the Schr\"odinger equation, 
constructed solely from the Hamiltonian function, and has shown that his 
equation leads to state vector reduction to an energy eigenstate, with   
energy conservation in the mean throughout the reduction process.    
In the simplest spin-1/2 case, Hughston exhibits an explicit solution 
that shows that his equation leads to collapse with the correct quantum 
mechanical probabilities, but the issue of collapse probabilities in the 
general case has remained open.   

Our purpose in this paper is to further investigate the structure 
and properties of Hughston's equation, proceeding from new identities in   
projective Hilbert space derived in Sec.~II.  
A principal result will be the proof in Sec.~III (using the martingale or 
``gambler's ruin'' argument pioneered by Pearle [7]) that 
in the generic case, with no approximations, Hughston's equation leads 
to state vector collapse to energy eigenstates with the 
correct quantum mechanical probabilities.  The relation of Hughston's 
equation to earlier work on norm-preserving equations is discussed in 
Sec.~IV, and the density matrix form of Hughston's equation, which gives 
a manifestly unitary stochastic evolution on pure states, is given in 
Sec.~V.  In Sec.~VI we examine the stochastic evolution of an initial 
state that is constructed as the product of independent subsystem states.  
Finally, in Sec.~VII we discuss whether an energy-based approach 
to stochastic evolution (as opposed to approaches [8] based on spontaneous 
localization) suffices to give a satisfactory objective description of 
the evolution of a state during the quantum mechanical measurement process.

\bigskip
\leftline{\bf II.~~PROJECTIVE HILBERT SPACE AND SOME IDENTITIES}
\bigskip

We begin by explaining the basic elements of projective Hilbert space 
needed to understand Hughston's 
equation, working in an $n+1$ dimensional Hilbert space.  We denote the 
general state vector in this space by $| z \rangle$, with $z$ a shorthand 
for the complex projections $z^0,z^1,...,z^n$ of the state vector on an 
arbitrary fixed basis.  Letting  $F$ be an arbitrary Hermitian operator, and 
using the summation convention that repeated indices are summed over their 
range, we define 
$$(F) \equiv { \langle z | F | z \rangle   \over \langle z |z \rangle } 
= { \overline z^{\alpha} F_{\alpha \beta} z^{\beta} \over  
\overline z^{\gamma} z^{\gamma} }~~~,
\eqno(1a)$$
so that $(F)$ is the expectation of the operator $F$ 
in the state $|z\rangle$, 
independent of the ray representative and normalization 
chosen for this state.  
Note that in this notation $(F^2)$ and $(F)^2$ are not the same; their 
difference is in fact the variance $[\Delta F]^2$, 
$$[\Delta F]^2 = (F^2)-(F)^2~~~.\eqno(1b)$$
We shall use two other parameterizations for the state $|z\rangle$ in what 
follows. Since $(F)$ is homogeneous of degree zero in both 
$z^{\alpha}$ and $\overline z^{\alpha}$, let us define new 
complex coordinates $t^j$ by  
$$t^j=z^j/z^0,~~ \overline t^j=\overline z^j 
/ \overline z^0~,~~~j=1,...,n, ~~~
\eqno(2)$$
which are well-defined over all states for which $z^0 \neq 0$ [9].  
Next, it is convenient to 
split each  of the complex numbers $t^j$ into its real and imaginary 
part $t^j_R,~t^j_I$, and to introduce a $2n$ component real vector 
$x^a,~a=1,...,2n$ defined by $x^1=t^1_R,~x^2=t^1_I,~x^3=t^2_R,~
x^4=t^2_I,...,x^{2n-1}=t^n_R,~x^{2n}=t^n_I$.   Clearly, specifying 
the projective coordinates $t^j$ or $x^a$ uniquely determines the 
unit ray containing the unnormalized state $|z\rangle$, while leaving 
the normalization and ray representative of the state $|z\rangle$ 
unspecified.   

As discussed in Refs. [6], projective Hilbert space is also a Riemannian  
space with respect to the Fubini-Study metric $g_{\alpha \beta}$, defined 
by the line element 
$$ds^2= g_{\alpha \beta} d\overline z^{\alpha} dz^{\beta}
\equiv 4\left( 1- { | \langle z | z+dz \rangle |^2 \over \langle z |z \rangle 
\langle z+dz | z+dz \rangle } \right) ~~~.\eqno(3a)$$ 
Abbreviating $\overline z^{\gamma} z^{\gamma} \equiv  \overline z \cdot z$, 
a simple calculation gives 
$$g_{\alpha \beta}=4(\delta_{\alpha \beta} \overline z \cdot z
-z^{\alpha} \overline z^{\beta})/(\overline z \cdot z)^2
=4 {\partial \over \partial \overline z^{\alpha} }
{\partial \over \partial z^{\beta} } \log \overline z \cdot z~~~.
\eqno(3b)$$
Because of the homogeneity conditions $\overline z^{\alpha} g_{\alpha \beta} 
=z^{\beta} g_{\alpha \beta}=0$, the metric $g_{\alpha \beta}$ is not 
invertible, but if we hold the coordinates $\overline z^0,~z^0$ fixed in  
the variation contained in  Eq.~(3a) and go over to the 
projective coordinates $t^j$, we can rewrite the line element of Eq.~(3a) 
as 
$$ds^2=g_{jk}d\overline t^j dt^k~~~,\eqno(4a)$$ 
with the invertible metric [9] 
$$g_{jk}={4[(1+\overline t^{\ell} t^{\ell}) \delta_{jk} - t^j \overline t^k ]
\over (1+\overline t^m t^m)^2 }~~~,\eqno(4b)$$                                 
                           
with inverse 
$$g^{jk}={1 \over 4} (1+\overline t^m t^m) (\delta_{jk} + t^j \overline t^k)
~~~.\eqno(4c)$$
Reexpressing the complex projective coordinates $t^j$ in terms of the 
real coordinates $x^a$, the line element can be rewritten as 
$$\eqalign{ 
ds^2=&g_{ab}dx^adx^b~~~,\cr
g_{ab}=&{4[(1+x^dx^d)\delta_{ab}-(x^ax^b+\omega_{ac}x^c\omega_{bd}x^d)] 
\over (1+x^e x^e)^2}~~~,\cr
g^{ab}=&{1\over 4} (1+x^e x^e)(\delta_{ab}+
x^ax^b+\omega_{ac}x^c\omega_{bd}x^d)~~~.\cr 
}\eqno(4d)$$
Here $\omega_{ab}$ is a numerical tensor whose only nonvanishing elements are  

$\omega_{a=2j-1 ~b=2j}=1$ and $\omega_{a=2j~b=2j-1}=-1$
for $j=1,...,n$.  As discussed 
by Hughston, one can define a complex structure $J_a^{~b}$ over the entire 
projective Hilbert space for which $J_a^{~c}J_b^{~d}g_{cd}=g_{ab},$   
$J_a^{~b}J_b^{~c}=-\delta_a^c$, 
such that $\Omega_{ab}=g_{bc} J_a^{~c}$ and 
$\Omega^{ab}=g^{ac}J_c^{~b}$ are antisymmetric tensors.  At $x=0$, the metric 
and complex structure take the values 
$$\eqalign{
g_{ab}=&4 \delta_{ab}~,~~g^{ab}={1 \over 4} \delta_{ab}~~~,\cr
J_a^{~b}=&\omega_{ab}~,~~\Omega_{ab}=4\omega_{ab}~,
~~\Omega^{ab}={1\over 4}\omega_{ab}~~~.\cr
}\eqno(5)$$

Returning to Eq.~(1a), we shall now derive some identities that are 
central to what follows.  Differentiating Eq.~(1a) with respect to 
$\overline z^{\alpha}$, with respect to $z^{\beta}$, and with respect to both 
$\overline z^{\alpha}$ and $z^{\beta}$, we get 
$$\eqalign{
\langle z | z \rangle {\partial (F) \over \partial \overline z^{\alpha}}
=&F_{\alpha \beta} z^{\beta} - (F) z^{\alpha}~~~,\cr
\langle z | z \rangle {\partial (F) \over \partial  z^{\beta}}=&
\overline z^{\alpha} F_{\alpha \beta}  - (F) \overline z^{\beta}~~~,\cr
\langle z | z \rangle^2 {\partial^2 (F) \over \partial \overline z^{\alpha} 
\partial z^{\beta} }=&
\langle z |z \rangle [F_{\alpha \beta}-\delta_{\alpha \beta} (F) ]
+2z^{\alpha} \overline z^{\beta} (F) - \overline z^{\gamma} F_{\gamma \beta} 
z^{\alpha}-\overline z^{\beta} F_{\alpha \gamma} z^{\gamma}~~~.\cr
}\eqno(6a)$$
Writing similar expressions for a second operator expectation $(G)$, 
contracting in various combinations with the relations of Eq.~(6a), and 
using the homogeneity conditions 
$$
\overline z^{\alpha} {\partial (F) \over \partial \overline z^{\alpha} }
=z^{\beta} {\partial (F) \over \partial z^{\beta} }
=\overline z^{\alpha} {\partial^2 (F) \over \partial \overline z^{\alpha} 
\partial z^{\beta}}
=z^{\beta} {\partial^2 (F) \over  \partial \overline z^{\alpha} 
\partial z^{\beta} } =0~~~\eqno(6b)$$
to eliminate derivatives with respect to $\overline z^0,~z^0$, we get 
the following identities,
$$\eqalign{
-i(FG-GF)=-i \langle z| z \rangle \left( 
{\partial (F) \over \partial z^{\alpha}} 
{\partial (G) \over \partial \overline z^{\alpha}} - 
{\partial (G) \over \partial z^{\alpha}} 
{\partial (F) \over \partial \overline z^{\alpha}} \right) 
=&2\Omega^{ab} \nabla_a (F) \nabla_b (G)~~~,\cr
(FG+GF)-2(F)(G)= \langle z| z \rangle \left( 
{\partial (F) \over \partial z^{\alpha}} 
{\partial (G) \over \partial \overline z^{\alpha}} + 
{\partial (G) \over \partial z^{\alpha}} 
{\partial (F) \over \partial \overline z^{\alpha}} \right) 
=&2g^{ab} \nabla_a (F) \nabla_b (G)~~~,\cr
(FGF)-(F^2)(G)-(F)(FG+GF)+2(F)^2(G)=\langle z | z \rangle ^2 
{\partial (F) \over \partial z^{\alpha}}
{\partial^2 G \over \partial \overline z^{\alpha} \partial z^{\beta}}
{\partial (F) \over \partial \overline z^{\beta}}
=&2\nabla^a (F) \nabla^b (F) \nabla_a \nabla_b (G),\cr
}\eqno(7a)$$
with $\nabla_a$ the covariant derivative constructed using the Fubini-Study 
metric affine connection.  It is not necessary to use 
the detailed form of this affine connection 
to verify the right hand equalities in these identities, because since $(G)$ 
is a Riemannian scalar, $\nabla_a \nabla_b (G)$$ =\nabla_a \partial_b (G)$, 
and since projective Hilbert space is a homogeneous manifold, it suffices 
to verify the identities at the single point $x=0$, where the affine 
connection vanishes and thus 
$\nabla_a \nabla_b (G)=\partial_a \partial_b (G)$.
Using Eqs.~(7a) and the chain rule we also find 
$$-\nabla_a [(F^2)-(F)^2] \nabla^a (G)=
-{1\over 2} (F^2 G+G F^2) +(F^2)(G) + (F) (FG+GF) -2(F)^2(G)~~~.
\eqno(7b)$$
When combined with the final identity in Eq.~(7a) this gives 
$$\eqalign{
D\equiv&\nabla^a(F) \nabla^b(F) \nabla_a \nabla_b (G)
-{1\over 2} \nabla_a [(F^2)-(F)^2] \nabla^a (G)\cr
=&{1\over 4}(2FGF-F^2G-GF^2)\cr
=&-{1\over 4}([F,[F,G]])~~~,\cr
}\eqno(7c)$$
with $[~,~]$ denoting the commutator, from which we see 
that $D$ vanishes when the operators $F$ and $G$ commute.  

An alternative derivation of Eq.~(7c) proceeds from the fact, noted by 
Hughston, that (for $F$ self-adjoint) 
$$\xi^a_F\equiv \Omega^{ab} \nabla_b(F)~~~\eqno(8a) $$ 
is a 
Killing vector obeying 
$$\nabla_c\xi_F^a+\nabla^a \xi_{cF}=0~~~.\eqno(8b)$$  
Using the identity $(F^2)-(F)^2=\nabla_b(F) \nabla^b(F)$, which is the 
$F=G$ case of the middle equality of Eq.~(7a), we rewrite $D$ of 
Eq.~(7c) as 
$$D=\nabla^a(F) \nabla^b(F) \nabla_a\nabla_b(G)
-\nabla^b(F)\nabla^a(G)\nabla_a \nabla_b(F)~~~.\eqno(9a)$$
This can be rewritten, using the identity 
$\Omega^{ab}\Omega_{cb}=\delta^a_c$, the antisymmetry of $\Omega$,  
the fact that $\Omega$ commutes with the covariant derivatives, 
and the Killing vector definition of Eq.~(8a), as 
$$\eqalign{
D=&\Omega^{ac} \xi_{cF} \Omega^{be} \xi_{eF} \nabla_a \nabla_b (G)
-\Omega^{bc} \xi_{cF} \Omega^{ae} \xi_{eG} \nabla_a \nabla_b (F)\cr
=&-\xi_{cF} \xi_{eF} \Omega^{ac} \nabla_a \xi^e_G
+\xi_{cF}\xi_{eG}\Omega^{ae}\nabla_a\xi^c_F~~~.
}\eqno(9b)$$
We now use the Killing vector identity of Eq.~(8b) on the final factor  
in each term, giving 
$$D=\xi_{cF} \xi_{eF} \Omega^{ac} \nabla^e \xi_{aG}
-\xi_{cF}\xi_{eG}\Omega^{ae}\nabla^c\xi_{aF}~~~.\eqno(9c)$$
Exchanging the labels $e$ and $c$ in the first term, and exchanging the 
labels $a$ and $e$ in the second term, we get 
$$\eqalign{
D=&\xi_{cF} \xi_{eF} \Omega^{ae} \nabla^c \xi_{aG}
+\xi_{cF}\xi_{aG}\Omega^{ae}\nabla^c\xi_{eF} \cr
=&\xi_{cF} \nabla^c[\Omega^{ae}\xi_{eF}\xi_{aG}]~~~.\cr  
}\eqno(10a)$$
Substituting the Killing vector definition of Eq.~(8a), this becomes   
$$\eqalign{
D=&\Omega_{cb}\nabla^b(F)\nabla^c[\Omega^{ae}
\Omega_{ef}\nabla^f(F)\Omega_{ag}\nabla^g(G)]\cr  
=&\Omega_{cb}\nabla^b(F)\nabla^c[\Omega_{gf}\nabla^g(G)\nabla^f(F)]\cr
=&-{1\over 4}([F,[F,G]])~~~,\cr
}\eqno(10b)$$
where to get the final line we have twice used the first identity in 
Eq.~(7a).   This completes our geometric derivation of Eq.~(7c). 
\bigskip  
\vfill\break
\leftline{\bf III.~~HUGHSTON'S EQUATION AND STATE VECTOR} 
\leftline{\bf ~~~~~~COLLAPSE PROBABILITIES}
\bigskip

Let us now turn to Hughston's stochastic differential equation, which 
reads  
$$dx^a=[2 \Omega^{ab}\nabla_b(H)-{1\over 4}\sigma^2 \nabla^aV]dt
+\sigma\nabla^a(H) dW_t~~~,\eqno(11a)$$
with $W_t$ a Brownian motion or Wiener process, with $\sigma$ a parameter 
governing the strength of the stochastic terms, with $H$ the Hamiltonian 
operator and $(H)$ its expectation, and with  $V$ the 
variance of the Hamiltonian, 
$$V=[\Delta H]^2=(H^2)-(H)^2~~~.\eqno(11b)$$
When the parameter $\sigma$ is zero, Eq.~(11a) is just [6] the transcription 
of the Schr\"odinger equation to projective 
Hilbert space.  For the time evolution of a 
general function $G[x]$, we get by 
Taylor expanding $G[x+dx]$ and using the It\^o stochastic calculus rules [10]
$$[dW_t]^2=dt~,~~[dt]^2=dtdW_t=0~~~,\eqno(12a)$$
the corresponding stochastic differential equation
$$dG[x]=\mu dt + \sigma \nabla_aG[x]\nabla^a(H) dW_t~~~,\eqno(12b)$$
with the drift term $\mu$ given by 
$$\mu=2 \Omega^{ab} \nabla_aG[x]\nabla_b(H)-{1\over 4}
\sigma^2\nabla^aV\nabla_a
G[x]+{1\over 2}\sigma^2 \nabla^a(H)\nabla^b(H)\nabla_a\nabla_bG[x]~~~.
\eqno(12c)$$
Hughston shows that with the $\sigma^2$ part of the drift term 
chosen as in Eq.~(11a), the drift term $\mu$ in Eq.~(12b) vanishes for the 
special case $G[x]=(H)$, guaranteeing conservation 
of the expectation of the energy with respect to the stochastic evolution 
of Eq.~(11a).  But referring to Eq. (7c) and the first identity in Eq.~(7a),  
we see that in fact 
a much stronger result is also true, namely that $\mu$ vanishes [and thus
the stochastic process of Eq.~(12b) is a martingale] whenever 
$G[x]=(G)$, with $G$ any operator that commutes with the Hamiltonian $H$.   

Let us now make two applications of this fact.  First, taking $G[x]=V=
(H^2)-(H)^2$, we see that the contribution from $(H^2)$ to $\mu$ 
vanishes, so the drift term comes entirely from $-(H)^2$.  
Substituted this into $\mu$ gives $-2(H)$ times the drift term produced by 
$(H)$, which is again zero, plus an extra term 
$$-\sigma^2 \nabla^a(H)\nabla ^b(H)\nabla_a(H)\nabla_b(H)
=-\sigma^2V^2~~~,\eqno(13a)$$
where we have used the relation $V=\nabla_a(H)\nabla^a(H)$ which follows 
from the $F=G=H$ case of the middle identity of Eq.~(7a).  Thus the 
variance $V$ of the Hamiltonian satisfies the stochastic differential 
equation, derived by Hughston by a more complicated method, 
$$dV=-\sigma^2 V^2 dt + \sigma \nabla_aV\nabla^a(H) dW_t~~~.\eqno(13b)$$  
This implies that the expectation $E[V]$ with respect to the stochastic 
process obeys 
$$E[V(t)]=E[V(0)]-\sigma^2 \int_0^t ds E[V(s)^2]~~~,\eqno(13c)$$
which using the inequality $0\leq E[\{V-E[V]\}^2]=E[V^2]-E[V]^2$ 
gives the inequality 
$$E[V(t)] \leq E[V(0)] -\sigma^2 \int_0^t ds E[V(s)]^2~~~.\eqno(13d)$$
Since $V$ is necessarily positive, Eq.~(13d) implies that $E[V(\infty)]=0$, 
and again using positivity of $V$ this implies that $V(s)$ 
vanishes as $s \to \infty$, apart from a set of outcomes 
of probability measure zero.  Thus, as concluded by 
Hughston, the stochastic term in his equation drives the system, as $t \to 
\infty$, to an energy eigenstate.  

As our second application of the vanishing of the drift term $\mu$ for 
expectations of operators that commute with $H$, let us consider the 
projectors $\Pi_e\equiv |e\rangle \langle e| $ on a complete set of 
energy eigenstates $|e \rangle$.  By definition, these projectors all 
commute with H, and so the drift term $\mu$ vanishes in the stochastic 
differential equation for $G[x]=(\Pi_e)$, and consequently the expectations   
$E[(\Pi_e)]$ are time independent; additionally, by completeness of   
the states $|e\rangle$, we have $\sum_e (\Pi_e)=1$.  
But these are just the conditions for   
Pearle's [7] gambler's ruin argument to apply.  At time zero, 
$E[(\Pi_e)]=(\Pi_e)\equiv p_e$
is the absolute value squared of the quantum mechanical amplitude  
to find the initial state in energy eigenstate $|e \rangle$.  At $t=\infty$, 
the system always evolves to an energy eigenstate, with the eigenstate  
$|f\rangle $ occurring with some probability $P_f$.  The expectation 
$E[(\Pi_e)]$, evaluated at infinite time, is then  
$$E[(\Pi_e)]=1 \times P_e + \sum_{f \neq e} 0 \times P_f = P_e~~~;\eqno(14)$$
hence $p_e=P_e$ for each  $e$ and the state collapses into energy eigenstates 
at $t=\infty$ with probabilities given by the usual quantum mechanical 
rule applied to the initial wave function.  

This conclusion clearly 
generalizes to the stochastic equation 
$$dx^a=[2 \Omega^{ab}\nabla_b(H)-{1\over 4}\sigma^2\sum_j \nabla^aV_j]dt
+\sigma\sum_j\nabla^a(H_j) dW_t^j~~~,\eqno(15a)$$
with the $H_j$ a set of mutually commuting self-adjoint 
operators that commute 
with $H$, with $V_j=(H_j^2)-(H_j)^2$, and with the $dW_t^j$ independent 
Wiener processes obeying $dW_t^jdW_t^k=\delta^{jk}dt$.  
Following the same method used in obtaining Eq.~(13b), and defining  
$C_{kj}=\nabla_a(H_k)\nabla^a(H_j)
=(H_kH_j)-(H_k)(H_j)$, one finds
$$dV_k=-\sigma^2 \sum_j C_{kj}^2 dt + \sigma \nabla_aV_k
\sum_j \nabla^a(H_j) dW_t^j~~~,\eqno(15b)$$  
and therefore 
$$E[V_k(t)]=E[V_k(0)]-\sigma^2 \int_0^t ds \sum_j E[C_{kj}(s)^2]~~~.
\eqno(15c)$$
Since $E[C_{kj}^2] \geq E[C_{kj}]^2$, we have as before 
$$E[V_k(t)] \leq E[V_k(0)]-\sigma^2 \int_0^t ds \sum_j 
E[C_{kj}(s)]^2~~~,\eqno(15d)$$
which implies that each $E[C_{kj}(s)]$ approaches zero as 
$s \to \infty$.  
Hence for each $k,j$ we have at large times 
$$E[(H_kH_j)]-E[(H_k)(H_j)] \to 0  ~~~,\eqno(15e)$$  
and so there is an effective quantum decorrelation of commuting observables.  
Moreover, for $k=j$ Eq.~(15e) implies that at large times $E[V_k] \to 0$,   
which since $V_k$ is nonnegative implies that $V_k$ approaches zero apart 
from a set of outcomes of probability measure zero,   
and so the state evolves to a simultaneous eigenstate of all the commuting 
observables entering the process of Eq.~(15a).
\bigskip
\leftline{\bf IV.~~RELATION OF HUGHSTON'S EQUATION TO OTHER}  
\leftline{\bf ~~~~ STOCHASTIC NORM-PRESERVING EQUATIONS}
\bigskip

Let us now specialize Eqs.~(12b) and (12c) to the case in which $G[x]$ is  
simply the expectation $(G)$ of an operator $G$.  Then by substituting  
Eqs.~(7c) and the second equality in Eq.~(7a), we find 
$$d(G)=\mu dt + \kappa dW_t~~~,\eqno(16a)$$
with 
$$\kappa={1\over 2}\sigma[(\{G,H\})-2(G)(H)]
={1\over 2}\sigma(\{G,H-(H)\})~~~,\eqno(16b)$$ 
and with
$$\mu=(-i[G,H])-{1\over 8}\sigma^2([H,[H,G]])~~~,\eqno(16c)$$
where we have used $\{~,~\}$ to denote the anticommutator.  

Let us now compare this with the evolution of $(G)$ implied by the 
stochastic state vector evolution 
$$d|z\rangle=[\alpha dt + \beta dW_t] |z\rangle ~~~,\eqno(17a)$$
with 
$$\eqalign{
\alpha=&-iH-{1\over 8}\sigma^2[A-(A)]^2~~~,\cr
\beta=&{1\over 2} \sigma [A-(A)]~~~,\cr
}\eqno(17b)$$
where $A$ is a general self-adjoint operator and $(A)$ is defined, as in 
Eq.~(1a), by $(A)=\langle z|A|z\rangle / \langle z|z \rangle$.  
For the evolution of 
$\langle z |G|z \rangle$, we find by the It\^o rules, 
$$\eqalign{
d\langle z |G|z \rangle =& \langle z | 
[\alpha^{\dagger} G + G \alpha+\beta^{\dagger} G \beta]dt 
+ [\beta^{\dagger}G+G \beta] dW_t |z\rangle\cr
=&\langle z|  -i[G,H]dt -{1\over 8}\sigma^2
[\{G,[A-(A)]^2\}-2[A-(A)]G[A-(A)]]dt\cr
&~~~~~~~~~+{1\over 2}\sigma \{G,[A-(A)]\} dW_t |z\rangle\cr  
=&\langle z| -i[G,H]dt -{1\over 8}\sigma^2[A,[A,G]]dt
+{1\over 2}\sigma \{G,[A-(A)]\} dW_t |z\rangle~~~.\cr  
}\eqno(18a)$$
When $G=1$, the right hand side of Eq.~(18a) vanishes, since the commutator 
terms vanish trivially and $\langle z|A-(A)|z \rangle=\langle z|A|z \rangle 
-\langle z|z \rangle (A)=0$.  Therefore the state vector evolution of 
Eqs.~(17a, b) is norm preserving, and so it is consistent to choose the   
normalization $\langle z|z \rangle =1$ in conjunction with this evolution.  
For general $G$  we then have $d\langle z |G|z \rangle=d(G)$, and 
so Eq.~(18a) gives   
an expression for $d(G)$, which we see is identical to Eqs.~(16a) - 
(16c) when the operator $A$ is taken as the Hamiltonian $H$.  In particular, 
when $A=G=H$ we learn from Eq.~(18a) that  
$d(H)=\sigma V dW_t$, in agreement with 
Eq.~(12b), because
$$\langle z| \{H,H-(H)\}|z\rangle = 2[(H^2)-(H)^2]=2V~~~,\eqno(18b)$$ 
and so the convergence argument of Eqs.~(13a)-(13d) follows directly 
from Eq.~(18a).  Apart from 
minor changes in notation, the norm preserving evolution of Eqs.~(17a) 
and (17b) is the one given by Gisin [2], Percival [3], and 
Ghirardi, Pearle, and Rimini [4], and so we see that this evolution is 
equivalent [11] to the state vector evolution in projective 
Hilbert space given by Hughston's equation.  

The evolution of Eq.~(17a) 
can be generalized [as was done for the Hughston equation in Eq.~(15)] 
to read
$$\eqalign{
d|z\rangle=&[\alpha dt +\sum_j \beta_j dW_t^j] |z\rangle~~~,\cr 
\alpha=&-iH-{1\over 8}\sigma^2\sum_j[A_j-(A_j)]^2~~~,\cr
\beta_j=&{1\over 2} \sigma [A_j-(A_j)]~~~,\cr
}\eqno(19)$$
with the $A_j$ any set of mutually commuting operators.  When the $A_j$ 
do not all commute with the Hamiltonian $H$, it is necessary to make the 
approximation of neglecting the Hamiltonian evolution (the $-iH$ term 
in $\alpha$) in proving that Eq.~(19) implies state vector reduction to 
the mutual eigenstates of the $A_j$ with probabilities given by the usual 
quantum mechanical rule.  Such a proof, very similar to the one given 
for Hughston's equation in Sec.~III above, has been given by 
Ghirardi, Pearle, and Rimini [4].  In order to carry through the proof 
with no approximations, it is necessary to assume that the $A_j$ are 
operators in the mutually commuting set $H_j$ that all commute with $H$, 
as was done in Sec. III.
\bigskip
\leftline{\bf V.~~DENSITY MATRIX EVOLUTION}

Let us now define the pure state density matrix $\rho$ by 
$$\rho={ |z\rangle \langle z| \over \langle z|z \rangle },~~~\eqno(20a)$$
in terms of which $(G)$ is given by 
$$(G)={\rm Tr} \rho G~~~.\eqno(20b)$$
Since $G$ is a fixed operator, Eq.~(18a) for $d(G)$ can be rewritten as  
$$\eqalign{
{\rm Tr} G d\rho =&{\rm Tr} \rho\left[ -i[G,H]dt 
-{1\over 8}\sigma^2[A,[A,G]]dt
+{1\over 2}\sigma \{G,[A-(A)]\} dW_t \right]\cr
=&{\rm Tr} G \left[ -i[H,\rho]dt   -{1\over 8}\sigma^2[A,[A,\rho]]dt
+{1\over 2}\sigma \{\rho,[A-(A)]\} dW_t \right]~~~,\cr
}\eqno(20c)$$
where in the final line we have cyclically permuted terms under the trace.  
Since Eq.~(20c) holds for arbitrary self-adjoint operators $G$, it implies 
that the density matrix obeys the stochastic differential equation (each 
term of which is self-adjoint) 
$$d\rho= -i[H,\rho]dt   -{1\over 8}\sigma^2[A,[A,\rho]]dt
+{1\over 2}\sigma \{\rho,[A-(A)]\} dW_t ~~~.\eqno(21a)$$
This equation can be written in an alternative form by observing that 
since $\rho$ is a pure state density matrix obeying $\rho^2=\rho$, we 
have $\rho(A)=\rho  {\rm Tr}\rho A=\rho A \rho$.  These facts imply that  
$$\eqalign{
\{\rho,[A-(A)]\}=& \rho A+A \rho -2 \rho(A)\cr
=&\rho^2 A + A \rho^2 - 2 \rho A \rho = [\rho,[\rho,A]~~~,\cr
}\eqno(21b)$$
and so we can rewrite Eq.~(21a) as  
$$d\rho= -i[H,\rho]dt   -{1\over 8}\sigma^2[A,[A,\rho]]dt         
+{1\over 2}\sigma [\rho,[\rho,A]] dW_t~~~. \eqno(21c)$$

Equations (21a) and (21c) have the following properties for 
general $A$:\hfill\break
\item{(i)}  Since ${\rm Tr} d\rho=0$, the condition ${\rm Tr} \rho=1$
is preserved by the time evolution.  
\item{(ii)}  After some algebra using the It\^o rules, one finds 
that $\rho^2=\rho$ implies that 
$$\{\rho,d\rho\}+[d\rho]^2=d\rho~~~,\eqno(21d)$$ 
which can be rewritten as $[\rho+d\rho]^2=\rho+d\rho$. Hence the evolution  
of Eqs.~(21a, c) is consistent with the pure state condition.  This is  
required by the fact that Eqs.~(21a, c) may be derived as consequences 
of Eqs.~(17a, b), which are a norm preserving pure state evolution.    
The condition of Eq.~(21d) determines the coefficient of the $\sigma^2...dt$ 
drift term in terms of the coefficient of the $\sigma...dW_t$ stochastic
term, and so the ratio of these two coefficients in Eq.~(21c) cannot be 
treated as an additional adjustable parameter.  
\item{(iii)}  Since $d\rho=d\rho^{\dagger}$, the self-adjointness of $\rho$ 
is preserved by the time evolution.  
\item{(iv)}  Time reversal invariance is violated by the stochastic terms, 
since when $dt$ and $i$ are reversed in sign,  the term $ -i[H,\rho]dt$ is  
invariant, but the term $-{1\over 8}\sigma^2[A,[A,\rho]]dt$ reverses sign. 
\item{(v)}  When we take the stochastic expectation of Eq.~(21c), the 
$dW_t$ term drops out, and we get 
$${dE[\rho] \over dt}=-i[H,E[\rho]] -{1\over 8}\sigma^2[A,[A,E[\rho]]]~~~, 
\eqno(22a)$$
which as pointed out by Percival [3] and Ghirardi, Pearle, and Rimini [4] 
is a quantum dynamical semigroup evolution of the completely 
positive Lindblad [12] form.  The stochastic expectation $E[\rho]$  is 
what is usually termed the density matrix; it starts off at $t=0$ as a 
pure state density matrix but then evolves, through the stochastic 
process, into a mixed state density matrix.  
\item{(vi)}  The conditions for Eq.~(22a) 
to admit stationary solutions $E[\rho]_S$ with $dE[\rho]_S/dt=0$ are very 
stringent, since when the left hand side of Eq.~(22a) is zero, multiplying 
by $E[\rho]_S$ and taking the trace gives 
$$0=-i {\rm Tr} E[\rho]_S [H,E[\rho]_S] 
-{1\over 8} \sigma^2 {\rm Tr} E[\rho]_S [A,[A,E[\rho]_S]] ~~~.\eqno(22b)$$
Using cyclic permutation under the trace, the first term on the right hand 
side vanishes, while the second term becomes 
$${1\over 8} \sigma^2 {\rm Tr} [A,E[\rho]_S]^2~~~,\eqno(22c)$$
which can only vanish when $[A,E[\rho]_S]=0$.  Substituting this 
equation back into Eq.~(22a) then further implies that $[H,E[\rho]_S]=0$.  
When $A$ and $H$ commute, these conditions can be satisfied with  
$E[\rho]_S$ a general function of $H$ [see the further discussion of 
this case in (xiii) below] , but when $A$ and $H$ do not 
commute,  one can have situations in which either $E[\rho]_S$  must 
be a multiple of the unit operator, which trivially commutes with both 
$H$ and $A$, or else there are no stationary 
solutions and $E[\rho]$ diverges at large times.  The latter case is  
found in spontaneous localization models, as discussed for 
example in Section III.B.3 of Ref. [4].  
\item{(vii)}  The evolution 
implied by  Eq.~(22a) leads [13] to a monotonic increase of the 
von Neumann entropy,  a result that can be demonstrated directly from  
Eq.(22a) as follows.   Letting 
$$S=-{\rm Tr} E[\rho] \log E[\rho]~~~\eqno(22d)$$
be the von Neumann  (or information) entropy, we find by 
substituting Eq.~(22a) and   
using cyclical permutation of factors under the trace, that 
$$\eqalign{
{dS \over dt}=& -{\rm Tr} {dE[\rho] \over dt} [1+\log E[\rho] ] \cr
=&{1\over 8} \sigma^2 {\rm Tr}[\log E[\rho],A][A,E[\rho]]~~~.\cr
}\eqno(22e)$$
Substituting  complete sets of eigenstates $|n\rangle,~|m\rangle$ 
of the nonnegative density 
matrix $E[\rho]$, this becomes 
$$\eqalign{
{dS \over dt}=&{1 \over 8} \sigma^2 \sum_n\sum_m 
[\log E[\rho]_n - \log E[\rho]_m]A_{nm} A_{mn} [E[\rho]_n-E[\rho]_m]  \cr
=&{1 \over 8} \sigma^2 \sum_n\sum_m 
|A_{nm}|^2 [\log E[\rho]_n - \log E[\rho]_m] [E[\rho]_n-E[\rho]_m] \geq 0
~~~.\cr
}\eqno(22f)$$
\item{(viii)}  Since Eqs.~(21a) and (21c) are nonlinear in $\rho$, the 
Schr\"odinger dynamics described by them cannot be represented as an 
equivalent Heisenberg or dual dynamics on the operator $G$.  On the other 
hand, Eq.~(22a) is linear in $\rho$, and so as noted by Lindblad, the 
Schr\"odinger dynamics for $E[\rho]$ can be represented as a dual Heisenberg 
dynamics for $E[G]$, given by 
$${dE[G] \over dt}=i[H,E[G]]-{1\over 8}\sigma^2[A,[A,E[G]]]~~~. 
\eqno(22g)$$
\item{(ix)}  The evolution of Eq.~(21c) can be written (after some 
algebra, and again using $\rho^2=\rho$) in the manifestly 
unitary form 
$$\rho+d\rho=U\rho U^{\dagger}~,~~~U=e^{dK}~~~,\eqno(23a)$$
with the infinitesimal anti-self-adjoint generator $dK$ given by 
$$dK=\left[ -iH-{1\over 8} \sigma^2[A^2-2A\rho A,\rho]\right] dt
-{1\over 2}\sigma[\rho,A]dW_t~~~.\eqno(23b)$$
Equations (21c) and (23b)  thus give the stochastic unitary extension of the 
Lindblad evolution of Eq.~(22a) [14].

Specializing to the Hughston case $A=H$, Eq.~(21c) (which uses the 
pure state condition $\rho^2=\rho$) becomes 
$$d\rho= -i[H,\rho]dt   -{1\over 8}\sigma^2[H,[H,\rho]]dt         
+{1\over 2}\sigma [\rho,[\rho,H]] dW_t~~~,\eqno(24a)$$
while Eq.~(22a) becomes 
$${dE[\rho] \over dt}=-i[H,E[\rho]] -{1\over 8}\sigma^2[H,[H,E[\rho]]]~~~, 
\eqno(24b)$$
and the following further properties are evident:\hfill\break
\item{(x)}  When $\rho=\Pi_e$, the projector on an energy eigenstate, 
then since all commutators in Eq.~(24a) vanish we have $d\rho=0$. 
\item{(xi)} For $G$ commuting with $H$, $E[d(G)]={\rm Tr }G dE[\rho]=0$, 
since by cyclic permutation inside the trace each term arising from 
substituting Eq.~(24b) into the expectation of Eq.~(20b) 
can be rearranged 
to have a factor $[G,H]$.  
\item{(xii)} For $V=[\Delta H]^2={\rm Tr} \rho H^2 - [{\rm Tr}\rho H]^2$, 
use of Eq.~(24a) and the It\^o calculus imply that $E[dV/dt]=-E[R^2]$, with 
$$R={1\over 2} \sigma {\rm Tr} [\rho,[\rho,H]] H =\sigma V~~~.\eqno(24c)$$
\item{(xiii)} Items (x) through (xii) are the ingredients used in Sec.~III 
to prove state vector collapse to energy eigenstates $|e\rangle$ 
with the correct quantum 
mechanical probabilities $p_e$. Hence at large times, as noted by 
Hughston, $E[\rho] \to \sum_e p_e \Pi_e$, which explicitly exhibits the  
role of $E[\rho]$ as the density matrix that evolves, under the stochastic 
process, from a pure to a mixed state form.  The fact that $E[\rho]$ for 
Hughston's equation approaches a stationary limit at large times 
is in accord with the general stationarity discussion given in (vi) above.  
\vfill\eject
\bigskip 
\leftline{\bf VI.~~BEHAVIOR OF SYSTEMS CONSTRUCTED} 
\leftline{\bf ~~~~~FROM INDEPENDENT SUBSYSTEMS}
\bigskip

Let us next examine the structure of Hughston's 
equation for a Hilbert space constructed as the direct product of  
independent subsystem Hilbert spaces, so that initially at time $t=0$ 
the state vector is 
$$|z\rangle = \prod_{\ell} |z_{\ell} \rangle~~~. \eqno(25a)$$   
We assume the Hamiltonian 
$$H=\sum_{\ell} H_{\ell}~~~,\eqno(25b)$$
with $H_{\ell}$ acting as the unit operator on the 
states $|z_{k}\rangle ~,~~
k \neq \ell$.  Then a simple calculation shows that the expectation 
of the Hamiltonian $(H)$ and its variance $V$  are both 
additive over the subsystem Hilbert spaces, 
$$\eqalign{
(H)=\sum_{\ell} (H_{\ell})_{\ell}~~~,\cr
V=\sum_{\ell} V_{\ell} =\sum_{\ell}[ (H_{\ell}^2)_{\ell} 
-(H_{\ell})_{\ell}^2]~~~,\cr
}\eqno(25c)$$
with $(F_{\ell})_{\ell}$ the expectation of the operator $F_{\ell}$ 
formed according to Eq.~(1a) with respect 
to the subsystem wave function $|z_{\ell}\rangle$.  In addition, 
the Fubini-Study line element is also additive over the subsystem Hilbert 
spaces, since 
$$\eqalign{
1-ds^2/4=& {| \langle z | z+dz \rangle |^2 \over \langle z |z \rangle 
\langle z+dz | z+dz \rangle } =\prod_{\ell}  
{ | \langle z_{\ell} | z_{\ell}+dz_{\ell} \rangle |^2 \over 
\langle z_{\ell} |z_{\ell} \rangle \langle z_{\ell}+dz_{\ell} 
| z_{\ell}+dz_{\ell} \rangle }\cr 
=&\prod_{\ell}[1-ds_{\ell}^2/4]=1-[\sum_{\ell} ds_{\ell}^2]/4 +{\rm O}(ds^4)
~~~.\cr}\eqno(26)$$
[An alternative way to see this is to use the identity 
$\log \overline z \cdot z =\log \prod_{\ell} \overline z_{\ell} 
\cdot z_{\ell} =$ $ \sum_{\ell} \log \overline z_{\ell}\cdot z_{\ell}$ 
in Eq.~(3b), along with a change of 
variable from $z$ to the $z_{\ell}$'s.]
As a result of Eq.~(26), the metric $g^{ab}$ and complex  
structure $\Omega^{ab}$ block diagonalize over the independent subsystem 
subspaces. Equations (25a)-(25c) then imply that 
Hughston's stochastic extension  
of the Schr\"odinger equation given in Eq.~(11a) 
separates into similar equations for the subsystems, that do not refer 
to one another's $x^a$ coordinates, but are correlated only through the  
common Wiener process $dW_t$ that appears in all of them.  These correlations 
result in the entanglement of the states $|z_{\ell}\rangle$, 
so that the product 
form of Eq.~(25a) is not maintained for times $t>0$, but subsystems 
$|z_{\ell}\rangle$ already in energy eigenstates 
remain unentangled for all time,  
since the coefficient of $dW_t$ vanishes in their stochastic evolution 
equations.  

These same conclusions follow from the density matrix form of Hughston's 
equation given in Eq.~(24a), in which the entanglements arising from the 
action of the same Wiener process on all subsystems are already evident, 
because the density matrix depends quadratically on the normalized state 
vector.  Considering for simplicity the case of two independent subsystems, 
substituting the $t=0$ form 
$$\rho = \rho_1 \rho_2 ~~~\eqno(27a)$$   
into Eq.~(24a), with $H=H_1+H_2$, we get
$$d\rho=d\rho_1 \rho_2+\rho_1 d\rho_2 
-{1 \over 4}\sigma^2 [H_1,\rho_1][H_2,\rho_2]dt  
~~~,\eqno(27b)$$
with $d\rho_1$ the evolution predicted by Eq.~(24a) within subsystem 1, 
$$d\rho_1= -i[H_1,\rho_1]dt   -{1\over 8}\sigma^2[H_1,[H_1,\rho]]dt         
+{1\over 2}\sigma [\rho_1,[\rho_1,H_1]] dW_t~~~,\eqno(27c)$$
and similarly for $d\rho_2$.  The entangling term proportional to 
$ [H_1,\rho_1][H_2,\rho_2]dt$ comes from the $[dW_t]^2$ contribution from   
the state vector evolution equation to the density matrix equation; it 
is in general nonzero, but vanishes when either $[H_1,\rho_1]=0$ or 
$[H_2,\rho_2]=0$, that is, when either of the two subsystems is in an 
energy eigenstate.  When more than two subsystems are present, the 
entangling term coupling $\rho_L$ to $\rho_{\ell}~,~~~\ell \neq L$ is 
more complicated in structure, but still has a factor  $[H_L,\rho_L]$
and so vanishes when the subsystem $L$ is in an energy eigenstate.  Thus 
the endpoints of the stochastic evolution under Hughston's equation, which 
are the energy eigenstates, can persist indefinitely as unentangled 
independent subsystems in a larger system.  

This conclusion does not extend to the more general 
evolution of Eq.~(21c), in which the stochastic process 
is driven by an operator $A$ differing from the Hamiltonian, with $A$ 
taken to be additive over subsystems.  The reason is that there is 
now a competition between the stochastic terms, which are constructed  
from double commutators with an innermost commutator $[A,\rho]$, and  
the Schr\"odinger evolution term, which involves the commutator $[H,\rho]$; 
the stochastic terms tend to drive the system to $A$ eigenstates, while 
the Schr\"odinger term  coherently mixes $A$ eigenstates, leading to   
evolution away from $A$ eigenstates.  Thus, a subsystem cannot 
remain indefinitely in an $A$ eigenstate, and 
as a result does not persist indefinitely 
as an unentangled independent subsystem in a larger system.  [These 
statements are in accord with the conclusions reached in the stationarity 
discussion of (vi) in Sec. V.]

\bigskip
\leftline{\bf VII.  DOES AN ENERGY-BASED EQUATION SUFFICE?}
\bigskip

In the preceding sections we have seen how Hughston's equation fits 
into the general framework of stochastic modifications of the 
Schr\"odinger equation that have been studied in the past.  Its 
distinguishing feature is that the general operator $A$ of Eqs.~(17a, b) and 
(21a, c) driving the                                                         
stochastic terms is chosen, in Hughston's case, to be the Hamiltonian $H$.  
This choice confers the advantage that the proof of reduction of the 
state vector to $A$ eigenstates (i.e., in Hughston's case, to energy  
eigenstates) with the correct quantum mechanical probabilities becomes 
exact, since it is not necessary to neglect the Hamiltonian evolution term.  
Moreover, since for Hughston's equation 
the stochastic expectation of the Hamiltonian operator 
$E[H]$ is conserved in time, and since convergence to $H$ eigenstates 
preserves the quantum mechanical predictions, 
any statistical test of energy conservation 
performed on the endpoint of the stochastic process will agree with the 
quantum mechanical prediction.  To justify these advantages, we must 
now address the issue of whether an energy-based stochastic equation is 
sufficient to give an objective interpretation of state 
vector reduction [15].  

First, we must deal with the objection that in most measurements, the 
quantum attribute being measured is not an energy; for example, in a 
Stern-Gerlach experiment, it is typically the $z$ component of a spin.  
However, to perform a measurement, it is always necessary to couple 
the quantum attribute being measured to the apparatus through an interaction 
energy term $H_I$, in such a way that the macroscopic state of the 
apparatus is ultimately determined by the quantum attribute being measured.  
Thus, in the first instance, what is being measured is an energy, even 
though after amplification to macroscopic scale this can be converted 
to other forms of indication, such as pointer displacements.  So from 
the point of view of the variety of quantum attributes that can be 
measured, Hughston's equation appears to be as viable as localizing 
approaches [8] in which $A$ is chosen as an operator that produces 
spatial localization.  

We must next deal with the issue of whether an energy-based approach can 
prevent the occurrence of macroscopic quantum superpositions.  For example, 
take a macroscopic object and displace it a macroscopic distance; the 
two states have the same energy, and so in Hughston's approach such 
superpositions would appear to be allowed, whereas in localizing approaches 
they are strongly forbidden.  However, this objection neglects the 
interactions of the macroscopic object with its environment, of the same  
type that are important in studies of decoherence.  When such effects are 
taken into account, macroscopic displacement of a macroscopic object 
results in an energy shift $\Delta E$, reflecting the altered environment,  
which is sufficient, from the 
point of view of Hughston's equation, to lead to rapid state vector 
reduction to one displaced alternative or the other.  To study this 
quantitatively, let us consider the following two environmental effects: 
(i) thermal energy fluctuations, and (ii) the surface adsorption of 
surrounding molecules.  Hughston proposes, as have other authors [16], that 
the  parameter governing the stochastic terms is of order 
$\sigma \sim M_{\rm Planck}^{-1/2}$ 
in microscopic units with $\hbar =c=1$, which he shows leads to   
state vector reduction in a time $t_R$ given by 
$$t_R \sim \left( { 2.8 {\rm MeV} \over \Delta E } \right)^2 {\rm sec}~~~.
\eqno (28)$$
Hence to get a reduction time of order, say, $10^{-6}$ seconds, one needs 
a  $\Delta E \sim 3 {\rm GeV} \sim 3~{\rm nucleon~masses}$.  

Considering first the effect of thermal fluctuations, let us consider 
a macroscopic object with $N\sim 10^{23}$ nucleon masses, 
so that $\Delta E \sim N^{1\over 2} kT \sim 8 {\rm GeV}$ at room temperature
($300^{\circ}$ Kelvin) and $\Delta E \sim .08 {\rm Gev}$ at the $3^{\circ}$  
temperature of the cosmic microwave background.  For such an object, 
thermal energy driven state vector reduction will occur in $10^{-7}$ seconds 
at room temperature and in $10^{-3}$ seconds at the temperature of the 
microwave background.  Examining next the effect of adsorbed molecules, 
consider an object with a surface area of $1~{\rm cm}^2$ at room temperature 
in an extreme vacuum of $10^{-14} {\rm Torr}$  (less [17] than the nighttime 
pressure at the surface of the moon.)   Then the flux of 
molecules bombarding its surface is [17] $4 \times 10^6$ per second, so 
assuming a high probability for the molecules to stick, 
a $\Delta E$ of $3 {\rm GeV}$ is attained in of order $10^{-6}$ seconds,    
permitting a $10^{-6}$ second state vector reduction time driven 
by the change in 
energy produced by surface adsorption.  One can scale to 
other sizes of macroscopic object from these examples, but they suffice 
to show that in the normal range of laboratory operating conditions for 
measuring apparatus, environmental interactions produce a large enough  
spread of energy values to give rapid state vector reduction through an 
energy driven stochastic equation.  

>From a formal point of view, it is instructive to cast the above discussion 
of environmental effects in terms of the analysis of the measurement 
process given by Zurek [18], starting from Eq.~(24b) for the evolution 
of the stochastic 
expectation of the density matrix. 
Zurek assumes that the total Hamiltonian $H$ describes the system $\cal S$ 
being measured, the apparatus $\cal  A$ doing the measuring, and the 
environment $\cal E$.  Thus, he writes the Hamiltonian as a sum of 6 terms, 
$$H=H_{\cal S}+H_{\cal A}+H_{\cal E}+H_{\cal SA}+H_{\cal AE}+H_{\cal SE}~~~,
\eqno(29)$$
with the first three terms giving the Hamiltonians of the system, apparatus, 
and environment in isolation from one another, and with the second three 
terms giving the corresponding interaction Hamiltonians.  Zurek assumes 
that the interaction $H_{\cal SE}$ between system and environment can be 
neglected, and that the interaction $H_{\cal SA}$ between system and 
apparatus acts only briefly while entanglement of the system and apparatus 
states is established, but is unimportant during the subsequent evolution 
of the density matrix that results in the actual measurement.  He also 
makes the simplifying assumption that the states which actually distinguish 
between quantities being measured have equal 
eigenvalues of the non-interaction part of the 
Hamiltonian $H_{\cal S}+H_{\cal A}+H_{\cal E}$, which    
implies that for the submatrix of $E[\rho]$ spanned by these states, the 
commutator $[H_{\cal S}+H_{\cal A}+H_{\cal E},E[\rho]]$ is zero, and so   
these commutator terms in Eq.~(24b) can be neglected.  With 
these simplifications, Eq.~(24b) becomes
$${dE[\rho] \over dt}=-i[H_{\cal AE},E[\rho]] 
-{1\over 8}\sigma^2[H_{\cal AE},[H_{\cal AE},E[\rho]]]~~~, 
\eqno(30a)$$
or when the non-Schr\"odinger term is omitted, as in Zurek's analysis, 
$${dE[\rho] \over dt}=-i[H_{\cal AE},E[\rho]] ~~~.\eqno(30b)$$
Zurek points out that the evolution of Eq.~(30b) introduces 
correlations between the apparatus and the environment, which select as 
the ``pointer basis'' of the apparatus, that registers the measurement,  
the eigenstates $|A_p\rangle$ of a ``pointer observable'' $\hat \Pi$ 
that commutes with $H_{\cal AE}$; in other words, the   
pointer basis projectors $\Pi_p=|A_p\rangle \langle A_p|$ must satisfy 
$$ [\Pi_p,H_{\cal AE}]=0~~~.\eqno(30c)$$
Returning to the full evolution equation of Eq.~(30a), 
with the non-Schr\"odinger terms included, we see that the 
argument of Sec.~III, 
when applied to this equation using Eq.~(30c), implies state vector 
collapse to the eigenstates of the Zurek pointer basis.    
Thus an energy-based stochastic reduction 
equation, when analyzed within the framework of Zurek's approximations, 
is consistent with, and adds further support to, 
the picture of the measurement process that 
Zurek proposes in [18].    

In addition to the issues just discussed, there are further  
questions that must be addressed  
in an energy-based approach, such as whether Hughston's estimated $\sigma$ 
gives sufficiently rapid (but also not too rapid) reduction of state 
vectors for all classes of experiments that have been carried out.  Answering 
this question is beyond the scope of the present paper, but is an 
important issue for future study.   Ultimately, the decision between an 
energy-based or localization-based approach (or yet some other choice of 
the operator $A$ driving the stochastic terms) may depend on which form 
of the modified Schr\"odinger equation can be derived as an  
approximation to relativistically invariant physics at a deeper level.  

To summarize, we have shown that Hughston's stochastic extension of the 
Schr\"odinger equation has properties that make it a viable physical model 
for state vector reduction.   This opens the challenge of seeing whether 
it can be derived as a phenomenological approximation to a fundamental 
pre-quantum dynamics, along the lines of existing work on open dynamical   
systems [19].  Specifically, we suggest that since 
Adler and Millard [20] have argued that quantum mechanics can emerge as 
the thermodynamics of an underlying non-commutative operator dynamics, 
and since the corrections to the thermodynamic approximation in this dynamics 
are driven by the trace of the energy operator 
multiplied by a coefficient parameter with dimensions of inverse mass, 
it may be possible to show that Hughston's stochastic process is the  
leading statistical fluctuation correction to this thermodynamics.

\bigskip

\centerline{\bf Acknowledgments}
This work was supported in part by the Department of Energy under
Grant \#DE--FG02--90ER40542.  One author (S.L.A.) wishes to thank J. 
Anandan for conversations introducing him to the Fubini-Study metric,    
F. Benatti for a conversation about evolutions of the Lindblad type, and 
G. C. Ghirardi and A. Bassi for emphasizing the relevance to our  
discussion of Ref.~[4] and for a stimulating discussion.   He also wishes 
to acknowledge the hospitality of the Aspen Center for Physics, where this 
manuscript was completed.  
The other author (L.P.H.) wishes 
to thank P. Leifer for many discussions on the properties of the complex 
projective manifold, and D. Moore for helpful conversations on this work.  
He is grateful to C. Piron of the University of Geneva, and the CERN Theory 
Division, for their hospitality during the final stages of this work.  
Helpful comments from referees are also acknowledged 
with appreciation.  

\vfill\eject
\centerline{\bf References}
\bigskip
\noindent
[1]  For a representative, but not exhaustive, survey of the earlier 
literature, see the papers of Di\'osi, Ghirardi et. al., Gisin, 
Pearle, and Percival cited by Hughston, Ref. [5] below.
\hfill\break
\medskip 
\noindent
[2]  N. Gisin, Helv. Phys. Acta {\bf 62}, 363 (1989).\hfill\break
\medskip
\noindent
[3]  I. C. Percival, Proc. R. Soc. Lond. A{\bf447}, 189 (1994).\hfill\break
\medskip
\noindent
[4]  G. C. Ghirardi, P. Pearle, and A. Rimini, Phys. Rev. A{\bf 42}, 
78 (1990).\hfill\break
\medskip
\noindent
[5]  L. P. Hughston, Proc. Roy. Soc. Lond. A {\bf 452}, 953 (1996).
\hfill\break
\medskip
\noindent
[6]  D. A. Page, Phys. Rev. A {\bf 36}, 3479 (1987); Y. Aharanov and 
J. Anandan, Phys. Rev. Lett. {\bf 58}, 1593 (1987); J. Anandan and Y. 
Aharanov, Phys. Rev. D {\bf 38}, 1863 (1988) 
and Phys. Rev. Lett. {\bf 65}, 1697 (1990); G. W. Gibbons, 
J. Geom. Phys. {\bf 8}, 147 (1992); L. P. Hughston, ``Geometric aspects 
of quantum mechanics'', in S. A. Huggett, ed., {\it Twistor theory}, 
Marcel Dekker, New York, 1995; A. Ashtekar and T. A. Schilling, preprint 
gr-qc/9706069.  For related work, see  
A. Heslot, Phys. Rev. D {\bf 31}, 1341 (1985) 
and S. Weinberg, Phys. Rev. Lett. {\bf 62}, 485 (1989) and 
Ann. Phys. (NY) {\bf 194}, 336 (1989).  \hfill\break
\medskip
\noindent
[7]  P. Pearle, Phys. Rev. D {\bf 13}, 857 (1976); Phys. Rev. D {\bf 29}, 
235 (1984); Phys. Rev. A {\bf 39}, 2277 (1989).\hfill\break  
\medskip
\noindent
[8]  See, e.g., Ref.~[4] above and G. C. Ghirardi, A. Rimini, and 
T. Weber, Phys. Rev. D {\bf 34}, 470 (1986).\hfill\break  
\medskip
\noindent
[9]  There are states which have some other element, say $z^k \neq 0$, 
and which overlap with those for which $z^0 \neq 0$. The function $(F)$,  
expressed in terms of the $t^j$'s, can then be extended by continuity 
to a function of a second set of variables $t^{j\prime} = z^j/z^k$, 
defined over the set $z^k \neq 0$.  With this process, one can extend  
the function $(F)$ to the covering projective space.
Similarly, in Eqs. (3a)-(4b), what we have called $z^0$ could 
be any $z^{\alpha}\neq 0$.  There is 
therefore a set of holomorphically overlapping patches, so that the 
metric of Eq.~(4b) is globally defined.  See, for example, S. Kobayashi 
and K. Nomizu, {\it Foundations of Differential Geometry}, Vol. II, p. 159, 
Wiley Interscience, New York, 1969. \hfill\break
\medskip
\noindent
[10]  For an excellent exposition of the It\^o calculus, see C. W. Gardiner, 
{\it Handbook of Stochastic Methods}, Springer-Verlag, Berlin, 1990, 
Chapt. 4.   \hfill\break
\medskip
\noindent
[11]  We wish to thank A. Bassi and G. C. Ghirardi for a conversation 
explaining this connection.  Our exposition closely follows theirs.\hfill  
\break
\medskip
\noindent
[12]  G. Lindblad, Commun. Math. Phys. {\bf 48}, 119 (1976); V. Gorini, 
A. Kossakowski, and E. C. G. Sudarshan, J. Math. Phys. {\bf 17}, 821 
(1976).  Abbreviating $\overline \rho \equiv E[\rho]$, the most general 
Lindblad-type evolution is 
$${d \overline \rho \over dt}= -i[H,\overline \rho]
+\sum_j[V_j\overline \rho V_j^{\dagger}- {1\over 2} V_j^{\dagger}V_j 
\overline \rho - {1\over 2} \overline \rho V_j^{\dagger}V_j]~~~,$$
which when $V_j^{\dagger}=V_j$ reduces to 
$${d \overline \rho \over dt}= -i[H,\overline \rho]
-{1\over 2}\sum_j[V_j,[V_j,\overline \rho]] ~~~,$$
corresponding to the structure of Eq.~(22a).  The positivity results 
of (vi) and (vii) below are special to the case of self-adjoint $V_j$, 
and do not extend to the general Lindblad-type evolution with 
$V_j^{\dagger} \neq V_j$. \hfill\break
\medskip
\noindent
[13]  For a general discussion, see G. Lindblad, {\it Non-Equilibrium 
Entropy and Irreversibility}, D. Reidel, Dordrecht, 1983, pp. 28-29; 
for a recent $2 \times 2$ matrix 
derivation and application to $K$-meson decays, see 
F. Benatti and R. Floreanini, in ``Quantum Probability'', 
Banach Center Publications, Vol. 43, Institute of Mathematics, 
Polish Academy of Sciences, Warsaw, 1998.\hfill\break  
\medskip
\noindent
[14]  We wish to thank F. Benatti for calling our attention to the 
book K. R. Parasarathy, 
{\it An Introduction to the Quantum Stochastic Calculus},  
Birkh\"auser Verlag, Basel, 1992, Chapt. III,
which discusses such extensions.\hfill\break
\medskip
\noindent
[15] For earlier discussions of energy-based reduction, see G. J. Milburn, 
Phys. Rev. A {\bf 44}, 5401 (1991); D. Bedford and D. Wang, Nuovo Cimento 
{\bf 26}B, 313 (1975) and Nuovo Cimento {\bf 37}B, 55 (1977).  \hfill\break
\medskip
\noindent
[16]  See L. P. Hughston, Ref. [5], Sec. 11 and earlier 
work of Di\'osi, Ghirardi 
et. al., and Penrose cited there; also D. I. Fivel, preprint 
quant-ph/9710042.\hfill\break  
\medskip
\noindent
[17]  P. A. Redhead, article on {\it Vacuum} in the Macmillan  
Encyclopedia of Physics, J. G. Rigden, ed., Simon \& Schuster Macmillan, 
New York, 1996, p. 1657.
\hfill\break
\medskip
\noindent
[18]  W. H. Zurek, Phys. Rev. D {\bf 24}, 1516 (1981).\hfill\break
\medskip
\noindent
[19]  See, e.g., H. Spohn, Rev. Mod. Phys. {\bf 53}, 569 (1980).\hfill\break
\medskip
\noindent
[20] S. L. Adler and A. C. Millard, Nucl. Phys. B {\bf 473}, 199 (1966); 
see also S. L. Adler and A. Kempf, J. Math. Phys. {\bf 39}, 5083 (1998).
\hfill\break
\bigskip
\noindent
\bye
\bigskip
\noindent
\bigskip
\noindent
\bigskip
\noindent
\bigskip
\noindent
\bigskip
\noindent
\bigskip
\noindent
\bigskip
\noindent
\bigskip
\noindent
\bigskip
\noindent
\bigskip
\noindent
\bigskip
\noindent
\bigskip
\noindent
\bigskip
\noindent
\bigskip
\noindent
\vfill
\eject
\bigskip
\bye